\documentclass[twocolumn]{aastex62}
\pdfoutput=1 
\usepackage{amsmath,amstext}
\usepackage[T1]{fontenc}

\def\kms{$\rm km~s^{-1}$}
\def\O3{O~III]}
\def\C4{C~IV}
\def\C3{$\rm C~III$ }
\def\HE2{He~II}
\def\N5{N~V}

\def\Si12{$\rm Si~XII$ }

\def\arcmin{$^\prime$}
\def\arcsec{\arcmin \arcmin}

\shorttitle{Cygnus Loop Shocks: I Shock Parameters}
\shortauthors{Raymond et al.}

\begin{document}

\title{Turbulence and Energetic Particles in Radiative Shock Waves in the Cygnus Loop I: Shock Properties}

\author[0000-0002-7868-1622]{John C. Raymond}
\email{jraymond@cfa.harvard.edu}
\affil{Center for Astrophysics | Harvard \& Smithsonian, 60 Garden St., Cambridge, MA 02138, USA}

\author[0000-0002-7924-3253]{Igor V. Chilingarian}
\affil{Center for Astrophysics | Harvard \& Smithsonian, 60 Garden St., Cambridge, MA 02138, USA}
\affil{Sternberg Astronomical Institute, M.V.Lomonosov Moscow State University, Universitetsky prospect 13, Moscow, 119234, Russia}

\author[0000-0003-2379-6518]{William P. Blair}
\affil{The Henry A. Rowland Department of Physics and Astronomy, Johns Hopkins University, 3400 N. Charles Street, Baltimore, MD 21218, USA}

\author[0000-0001-8858-1943]{Ravi Sankrit}
\affil{Space Telescope Science Institute, Baltimore, MD, USA}

\author[0000-0002-7597-6935]{Johnathan D. Slavin}
\affil{Center for Astrophysics | Harvard \& Smithsonian, 60 Garden St., Cambridge, MA 02138, USA}

\author[0000-0001-5817-5944]{Blakesley Burkhart}
\affil{Center for Computational Astrophysics, Flatiron Institute, 162 Fifth Ave., New York, NY 10010, USA}
\affil{Department of Physics and Astronomy, Rutgers, The State University of New Jersey, 136 Frelinghuysen Rd., Piscataway, NJ 08854, USA}

\date{Jan 30, 2020}


\begin{abstract}
We have obtained a contiguous set of long-slit spectra of a shock wave in the Cygnus Loop to investigate its structure, which is far from the morphology predicted by 1D models.  Proper motions from Hubble Space Telescope images combined with the known distance to the Cygnus Loop provide an accurate shock speed.  Earlier analyses of shock spectra estimated the shock speed, postshock density, temperature, and elemental abundances.  In this paper we determine several more shock parameters: a more accurate shock speed, ram pressure, density, compression ratio, dust destruction efficiency, magnetic field strength, and vorticity in the cooling region.  From the derived shock properties we estimate the emissivities of synchrotron emission in the radio and pion decay emission in the gamma rays. Both are consistent with the observations if we assume simple adiabatic compression of ambient cosmic rays as in the van der Laan mechanism.  We also find that, although the morphology is far from that predicted by 1D models and the line ratios vary dramatically from point to point, the average spectrum is matched reasonably well by 1D shock models with the shock speed derived from the measured proper motion.  A subsequent paper will analyze the development of turbulence in the cooling zone behind the shock.
\end{abstract}


\section{INTRODUCTION}

While some supernova remnant (SNR) shock waves seen in optical emission lines resemble 1D, steady-flow models of the cooling gas behind a shock \citep{cox72, raymond79, allen08},
those in the Cygnus Loop appear much different in higher temperature lines such as [O III] than in cooler lines such as H$\alpha$ or [S II].  The clumpy structure seen in the cooler lines suggests that turbulence develops as the gas cools.

Turbulence is ubiquitous in the interstellar medium (ISM), spanning scales from kilometers to hundreds of parsecs \citep{armstrong95, spangler01, burkhart10, krumholz16, chepurnov15}.  The turbulent cascade is generally believed to arise from the energy injected by supernova explosions, largely because the size scale and total momentum match, though it has also been suggested that an accretion-like flow and mass transport in the Galactic disk could inject energy \citep{krumholz18}.  The turbulence in the Warm Interstellar Medium (WIM) matches a Kolmogorov spectrum on average, though power can be enhanced locally \citep{burlaga18}.  Turbulence in molecular clouds is compressible, with high sonic Mach numbers \citep{burkhart13, federrath08}.   In the simplest interpretation of the Kolmogorov spectrum, energy is conserved as it cascades from large to small scales, and the persistence of the Kolmogorov spectral shape over many orders of magnitude would suggest that there is no substantial injection or dissipation of energy at intermediate scales.  Thus in effect the momentum injection should occur only as supernova remnants merge into the ISM and the speed of the shell declines to around 15 \kms .

Interstellar shock waves are classified as `radiative' or `nonradiative' depending on whether radiative cooling has had time to affect the dynamics of the postshock flow \citep{drainemckee}.  In the nonradiative shocks, the plasma properties remain more or less constant behind the shock, while in radiative shocks the gas cools catastrophically from the immediate postshock temperature to a recombination/photoionization layer at around 10,000 K.  It finally cools to temperatures below 1000 K, where it emits little radiation at visible wavelengths.  Typical SNR models indicate that shocks become radiative when the shock speed falls to about 400 \kms and the post-shock temperature drops below about $2 \times 10^6 ~\rm K$.

There are many analytic and numerical studies of the interaction of shock waves with turbulence \citep{bykov82, moczburkhart, robertson18}.  In the gasdynamic regime, the shock compresses and modifies the turbulence \citep{zank07}. Interaction of the shock with density inhomogeneities generates vorticity, which can amplify magnetic fields \citep{giacalone07, xulazarian, guo12}.  On smaller scales, collisionless shocks are inherently turbulent in the supercritical regime (Mach numbers above $\sim$2.7, where a steady flow cannot dissipate the energy), and they generate a variety of wave modes that can transfer energy among particle species, accelerate particles, and amplify magnetic fields \citep{blandfordeichler, bell04}.  These processes are important for the structure and evolution of the nonradiative shocks that dominate the X-ray emission of SNRs. 

As gas behind a shock cools, turbulence can be further amplified, but additional fluid instabilities can also arise.  The gas is subject to thermal instability for shock speeds above about 150 \kms \citep{gaetz88, innes92, sutherland03} and to a thin shell instability if a shell of cool, dense gas is driven by the pressure of hot interior gas \citep{vishniac83}.  In addition, for typical ISM conditions, the cooling gas will transition from gas pressure-dominated to magnetic pressure-dominated (high plasma $\beta$ to low $\beta$) conditions, so that fluctuations in magnetic field strength lead to density fluctuations \citep{raymondcuriel}.  Beyond that, vorticity generated at the shock continues to cascade and be amplified as the cooling gas is compressed.  It takes time for these instabilities to develop, and some SNRs, such as G65.3+5.7 \citep{mavromatakis02}, show radiative shocks that are remarkably smooth.  On the other hand, the Cygnus Loop and others show fluffy, fragmented structure that appears much different in lines formed at different temperatures.  

The Cygnus Loop is a middle-aged SNR about 21,000 years old and 42 pc in diameter \citep{fesen18}.  Nonradiative shocks in low-density gas produce X-ray emission in the north and south, but to the east and west the remnant drives 100--150 km/s radiative shocks into higher density clouds.  Stready-flow 1D models are generally able to match the UV, optical, and IR spectra of these radiative shocks \citep{raymond88, sankrit14, dopita16}, but the morphology of the emission changes drastically between the smooth, well-defined sinuous filaments seen in [O III] and the clumpy, more chaotic emission seen in H$\alpha$ and other cool lines.  That raises the questions of what determines the morphology of the cooler gas, how turbulence develops in the postshock flow, and whether the 1D models that match observed spectra can really be used to determine shock speeds and elemental abundances.

In order to quantify these effects and understand the physics of the postshock turbulent flow, we have acquired Hubble Space Telescope (HST) spectra and narrowband images, along with ground-based spectra covering a swath of the radiative shock wave in the western Cygnus Loop.  In this paper we focus on the ground-based spectra, supplemented by archival HST images that we use to measure proper motions.  In conjunction with the 735$\pm$25 pc distance \citep{fesen18}, they determine shock speed.  The spectra provide line intensities, line centroids, and gradients, from which we determine the basic parameters of the shock, including pre- and postshock densities and magnetic fields, ram pressure, compression ratio and vorticity.  We use these parameters to compute the intensities in radio and gamma-ray emission assuming simple adiabatic compression of the ambient cosmic-rays and magnetic field \citep{vanderlaan}.  The HST data and our conclusions about the development of turbulence in the postshock cooling region will be discussed in a separate paper.

\section{OBSERVATIONS AND DATA REDUCTION}
\begin{figure}
  \center
    \includegraphics[width=\hsize]{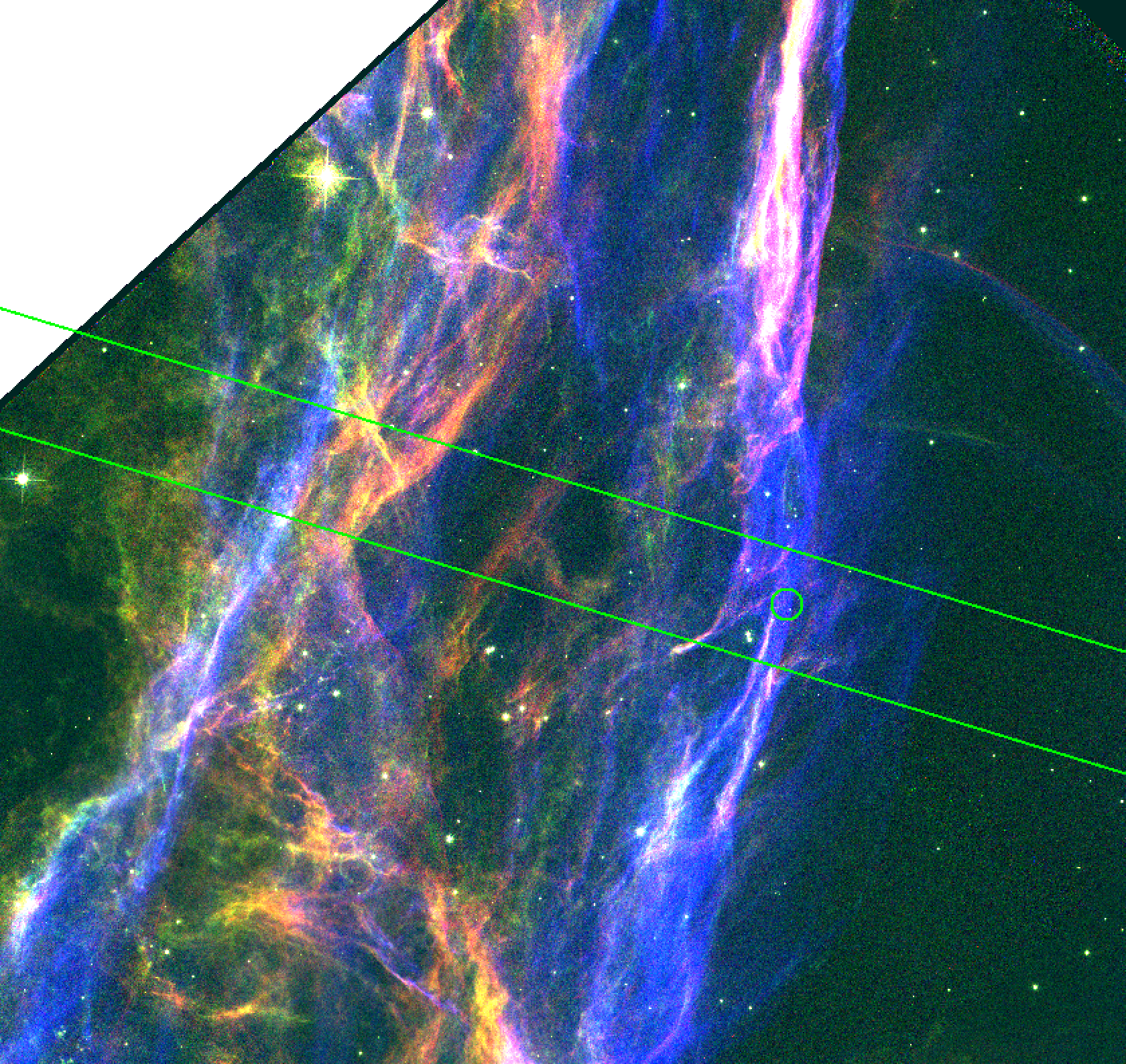}
\caption{Region covered by Binospec spectra overlaid on a composite HST image in the [O III] (blue), H$\alpha$ + [N II] (red) and [S II] (green) bands from the WFC3 camera (Hubble Heritage Project, http://heritage.stsci.edu/2015/29/index.html.  Twenty-three long-slit spectra were obtained between the green lines.  The shock is moving approximately along the green lines toward the lower right.  North is up and east is toward the left.  The separation between the green lines is 15\arcsec.  The circle indicates the 0\arcsec position in Figures 4, 5, 7--10, and 12, with the position coordinate increasing toward the right.
\label{overlay}
}
\end{figure}

We obtained spectra with the Binospec instrument, an efficient spectrograph designed for multi-object spectroscopy \citep{fabricant19} and operated at the 6.5-m converted MMT at Mt.~Hopkins, Arizona since November of 2017.  We used it with a single 0.75~arcsec wide 15~arcmin long slit in a ``quasi-IFU mode'', offsetting the slit center by 0\farcs75 perpendicular to the slit between exposures. The 23 exposures were acquired at a position angle of 73$^\circ$, and they covered a region centered at 20$^h$ 45$^m$ 37.6146$^s$, +31$^\circ$ 00$^\prime$ 09\farcs2.  An overlap of two exposures was used to ensure consistency between the two observing nights.  Figure~\ref{overlay} shows the region observed overlaid on a composite [O~III], H$\alpha$ + [N~II], and [S~II] image from the Hubble Heritage program (http://heritage.stsci.edu/2015/29/index.html). Binospec simultaneously obtained spectra of another section of the Cygnus Loop, which we will not discuss here. 

Observations were acquired on the nights of 2018 June 7 and June 11. Clouds interupted the sequence on June 7, so the width of the region observed was truncated and the last exposure was effectively exposed about half as long as the others.  The 600 l/mm grating was used with a slit width of 0\farcs75, giving a dispersion of 0.62 \AA\/ per pixel and a resolution of 1.5~\AA\ FWHM, which translates to resolving power $R\approx3200$ around H$\beta$ and $R\approx4400$ around H$\alpha$. The 4Kx4K E2V CCD has 0\farcs24 pixels that cover 9.6~arcmin along the slit, and the spectral range was 4600--7100 \AA .  We obtained 540 seconds of exposure time at each position split into three 180~s exposures.  The atmospheric seeing quality was about 1\farcs1, so there is effectively some correlation between neighboring slit positions. 

We used a special ``series'' mode of the Binospec data reduction pipeline \citep{2019PASP..131g5005K} to reduce all exposures in every quasi-IFU sequence at once. The pipeline performed primary reduction, cosmic-ray cleaning, flat-fielding, and wavelength and flux calibration. Wavelength calibration was accomplished with He and Ar lines from arc lamp spectra obtained immediately after the end of every sequence of science observations, and measurements of the bright night sky lines showed that centroids are accurate to 2--3 \kms. Flux calibration was based on pre-computed Binospec throughput measurements stored in the pipeline based on the standard star BD+17$^\circ$4807. It provided relative flux levels accurate to within 3--4~\%\ across the full spectral range.  However, the absolute fluxes are not guaranteed because of non-photometric conditions during observations, and we have scaled the fluxes by about 25\% to match the HST [O III] and H$\alpha$ images discussed in the next section.  We disabled the sky subtraction feature of the pipeline and subtracted sky background estimated from a section of the slit outside the Cygnus Loop from each exposure at the end of data reduction.

\begin{figure}
  \center
    \includegraphics[clip,trim={1.0in 5.0in 1.8in 0},width=\hsize]{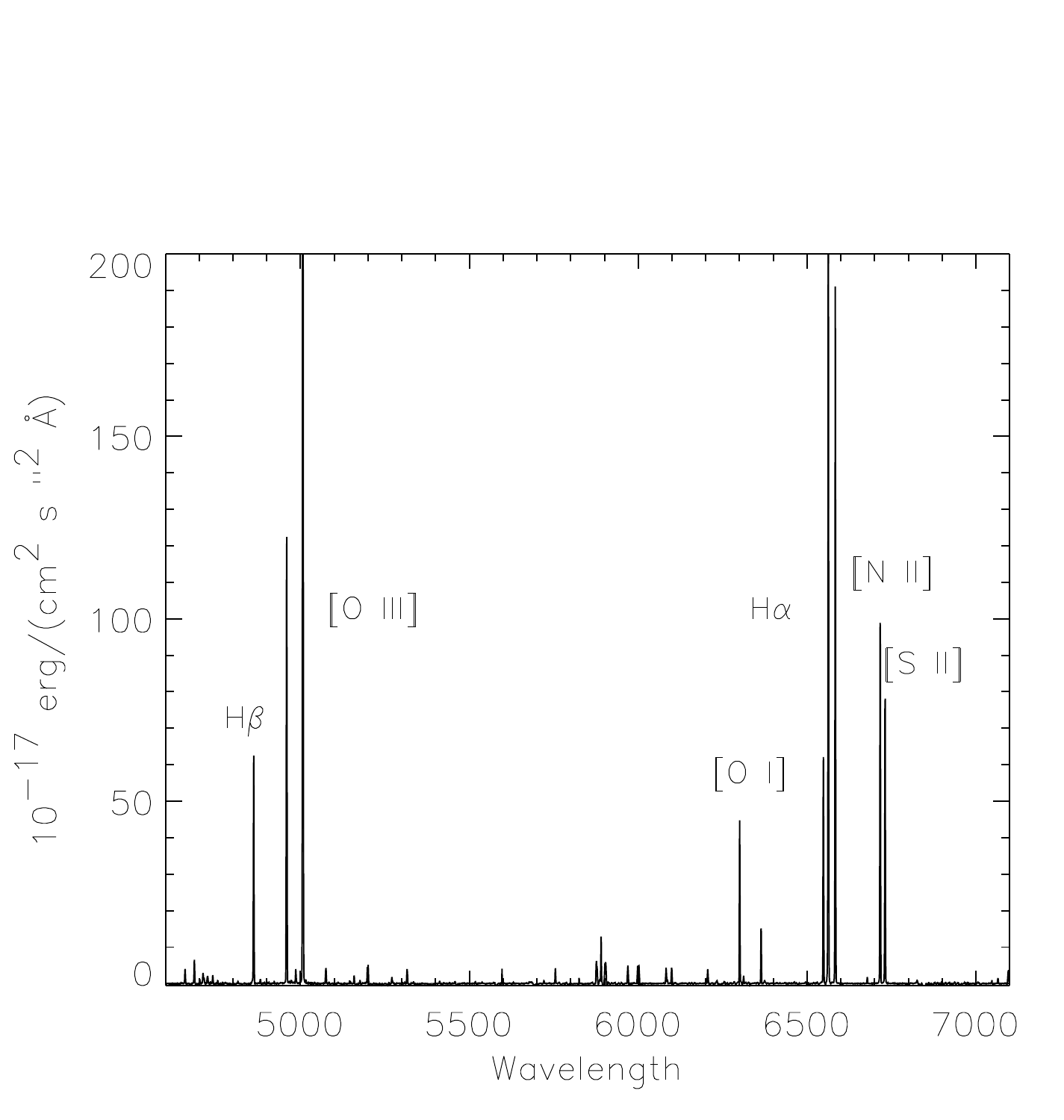}\\
    \includegraphics[clip,trim={1.0in 5.0in 1.8in 1.0in},width=\hsize]{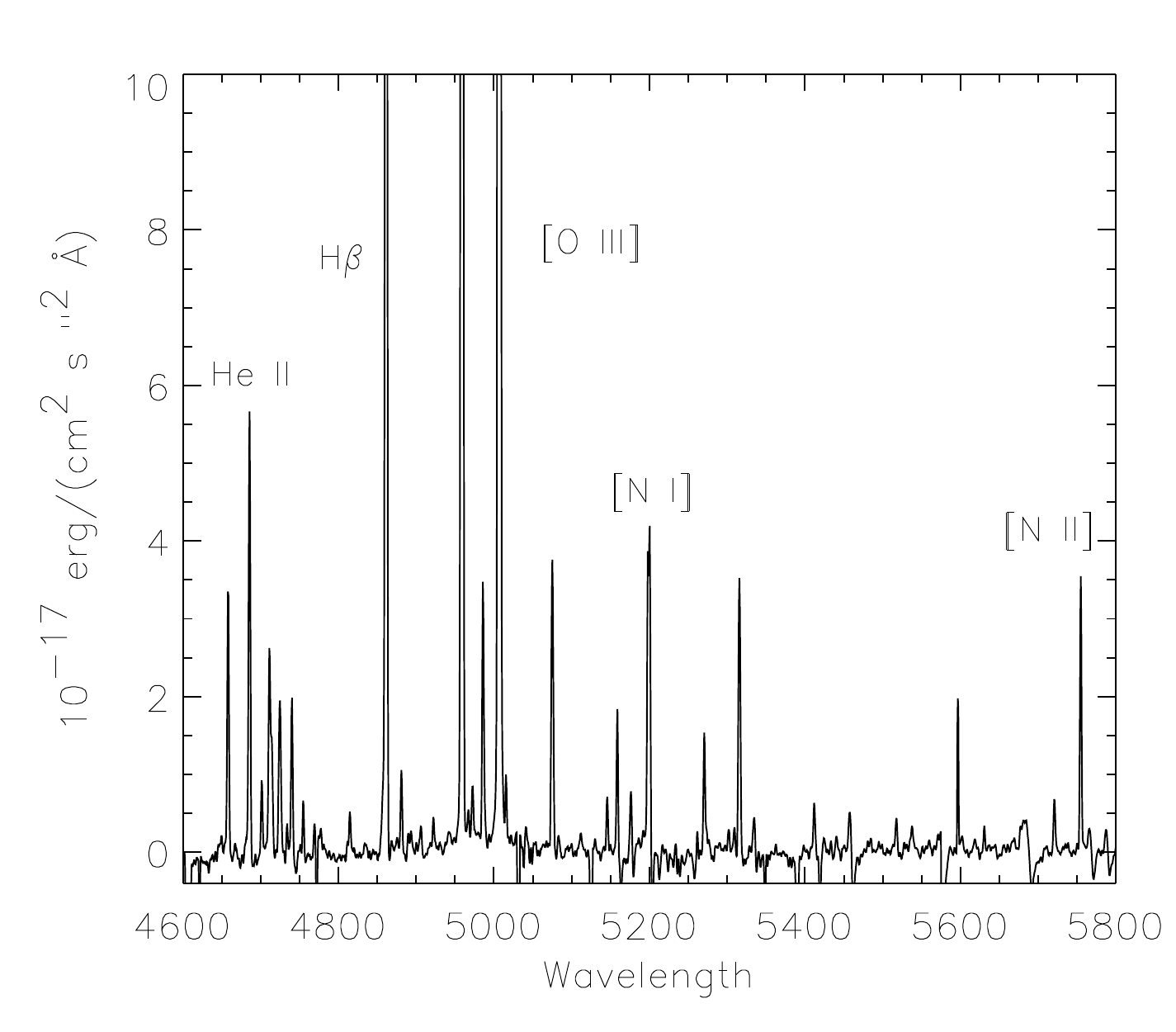}
\caption{Top: average of 10 spectra over the 110" where the emission is bright.  Negative values are imperfectly subtracted night sky lines. Bottom: blowup of the blue part of the spectrum showing weaker lines.  Negative values are imperfectly subtracted night sky lines.
\label{spct1}
}
\end{figure}

\begin{table}
\begin{center}
\caption {Emission line intensities}
\begin{small}
\begin{tabular}{l l c c c c}
\hline
\hline
Wavelength   &  Species   &  F$^a$    &$^b$   I   &  Mod.130  & Mod.130i \\
    \AA   &                   &                  &    & &    \\
\hline
 4658.54    & [Fe III]    & 6.9     & 7.0    & - & - \\
 4685.82    & He II       & 11.3     & 11.7   &      7.7  &        11.7 \\
 4701.32    & [Fe III]    & 1.8     & 1.8    & - & - \\
 4711.24    & [Ar IV]     & 5.6     & 5.7    &      2.6  &        4.1 \\
 4724.88    & [Ne IV]     & 2.5     & 2.5    &      1.5  &        3.6 \\
 4740.38    & [Ar IV]     & 3.9     & 3.9   &       1.9  &        3.0 \\
 4861.28    & H$\beta$    & 100.   &  100.  &      100  &      100 \\
 4922.04    & He I        & 0.9     & 0.9   &      1.0  &        0.8 \\
 4959.24    & [O III]     & 215.   & 213.   &  115  &        178 \\
 4972.88    & [Fe II]   . & 1.4     & 1.4   & - & - \\
 4985.90    & [Fe III]    & 7.0     & 7.0   & - & - \\
 5006.98    & [O III]    & 657.    & 650.   &      345  &        533 \\
 5015.66    & He I        & 1.8     & 1.8   &      2.2  &        1.7 \\
 5145.86    & [Fe VI]     & 1.7     & 1.7   &      4.4  &        6.8 \\
 5159.88    & [Fe II,VII] & 3.2     & 3.2   & - & - \\
 5176.24    & [Fe VI]     & 1.7     & 1.6   &    2.3 &        3.7 \\
 5200.42    & [N I]       & 14.0    & 12.9  &      11.2   &      7.7 \\
 5270.48    & [Fe III]    & 3.0     & 2.9   & - & -  \\
 5517.24    & [Cl III]    & 1.1     & 1.0   & - & - \\
 5630.70    & [Fe VI]     & 0.7     & 0.6   &       2.4  &        3.7 \\
 5720.60    & [Fe VII]    & 1.2     & 1.1   &       3.6  &        5.6 \\
 5754.70    & [N II]      & 7.0     & 6.1   &       6.0  &        5.8 \\
 5875.60    & He I        & 19.0    & 16.6  &       10.5  &       8.4 \\
 6086.40    & [Fe VII]    & 9.3    & 8.1    &       3.6   &       8.3 \\
 6300.30    & [O I]      & 76.0    & 63.8   &       29.2  &      21.5 \\
 6312.08    & [S III]    &  4.0     & 3.4   &       2.2   &       3.4 \\
 6364.16    & [O I]      & 25.3     & 21.1  &       9.7  &     7.2 \\
 6374.08    & [Fe X]     &  3.0     & 2.5   &  0 & 0 \\
 6548.30    & [N II]     & 110.    & 90.7   &       63    &      57 \\
 6563.18    & H$\alpha$  &  399.   & 329.   &      300    &     300 \\
 6583.64    & [N II]     & 338.    &279.    &      189   &      170 \\
 6678.50    & He I       &  3.2     & 2.6   &      3.0   &      2.4 \\
 6716.32    & [S II]     & 183.    &151.    &      157    &     162 \\
 6731.20    & [S II]     & 150.    &121.    &      118    &     132 \\
 7004.62    & [Ar V]     &  1.2     & 1.0  &      1.6   &      2.5 \\
 7065.38    &He I        & 3.2      &2.5    &     1.8    &     1.5 \\
\hline
H$\beta$    &            & 13.6$^c$      &  20.2$^c$ &  17.4$^c$    & 11.0$^c$  \\  
\hline
\end{tabular}
$^a$  Observed \\
$^b$  Corrected for E(B-V) = 0.2 \\
$^c$ H$\beta$ in units of $10^{-16}~\rm erg~cm^{-2}~s^{-1}~ arcsec ^{-2}$ \\
\end{small}
\end{center}
\label{spectrum}
\end{table}

Fig.~\ref{spct1} and Table~\ref{spectrum} show the average of 23 spectra covering a 110\arcsec \/ section of the slit after scaling the spectrum that was affected by clouds to match the others.  Negative values are imperfectly subtracted night sky lines.  Note the presence of faint lines of [Fe~VI], [Fe~VII] and [Fe~X] in Table 1.  Complementary HST UV and optical spectra of part of the observed region will be discussed in a subsequent paper.


\section{ANALYSIS}

\subsection{Proper Motion}

\begin{figure}
  \center
    \includegraphics[width=\hsize]{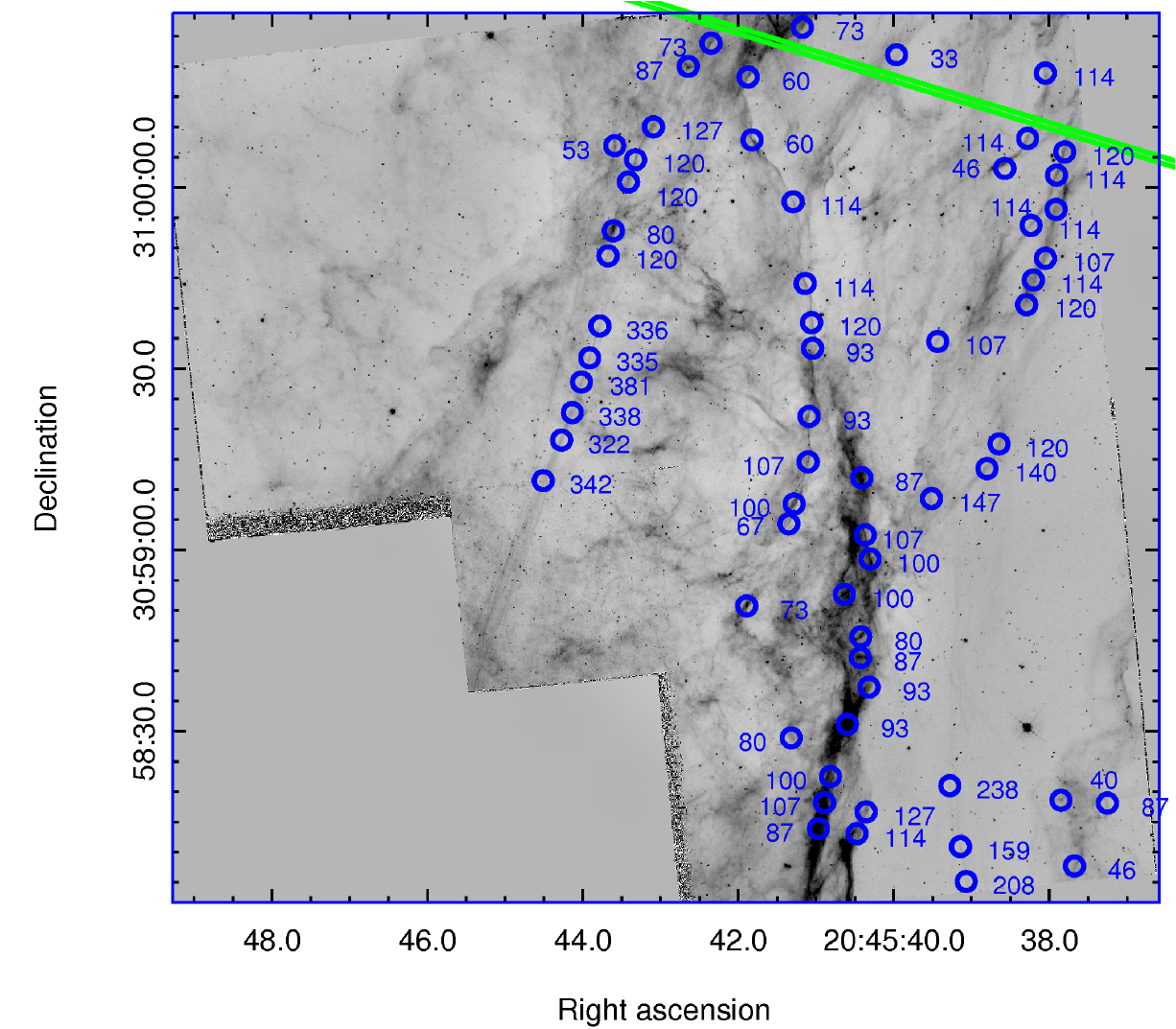}
    \includegraphics[width=\hsize]{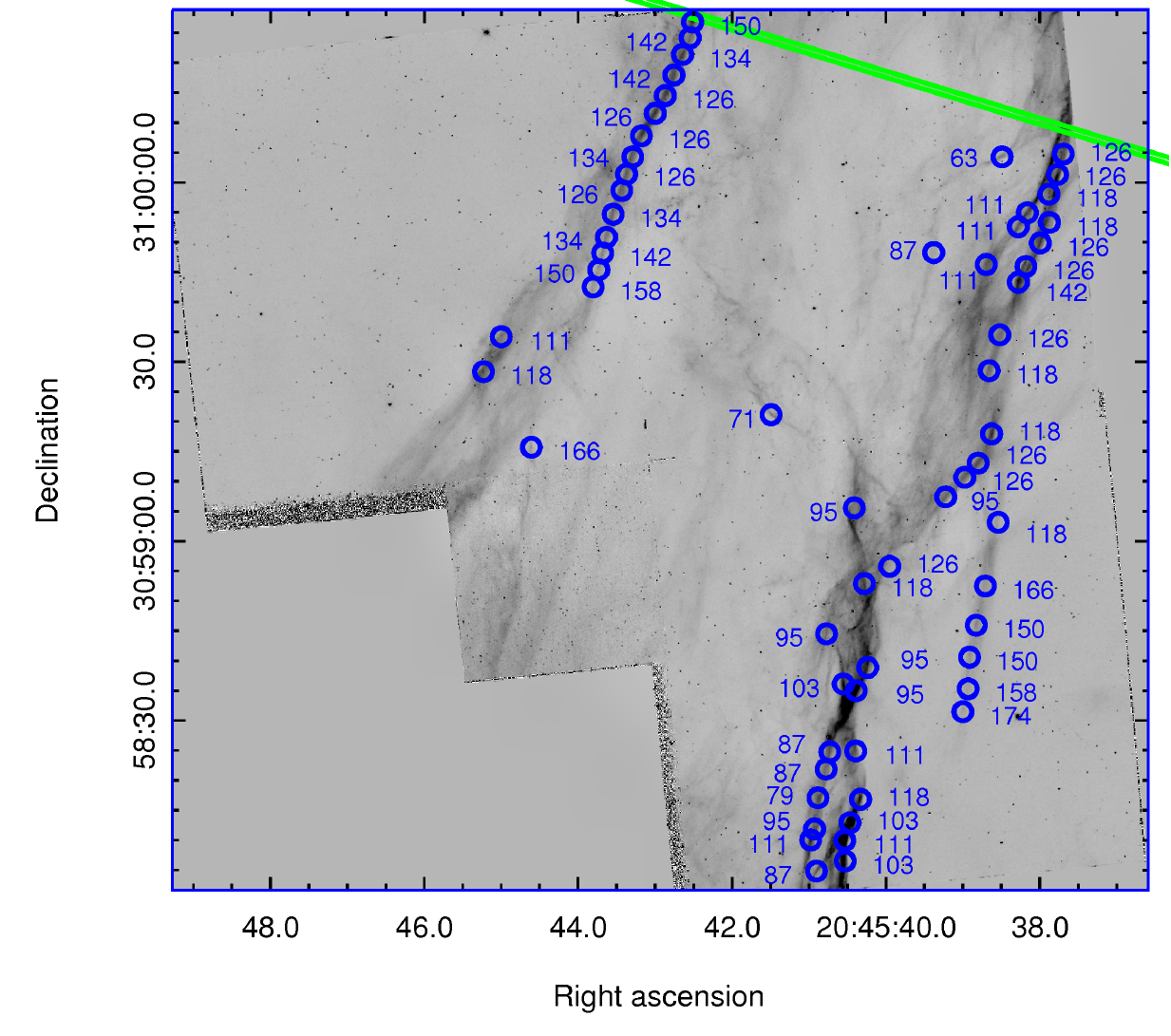}
\caption{H$\alpha$ (top) and [O III] (bottom) images from WFC2 on HST from 1997.  Circles indicate where proper motions were measured and numbers give the velocities in \kms assuming a distance of 735~pc \citep{fesen18}. The green lines indicate the slit position where the parameters shown in Figures 4--5, 7--10 and 12 were extracted.
\label{pm_ha}
}
\end{figure}

We compared HST WFPC2 images in H$\alpha$ and [O~III] obtained by J. Hester on 1997 August 17 (Observing Program 5779) with the [O III] image from the Hubble Heritage Project obtained on 2015 April 14-17 and an H$\alpha$ image that we obtained on 2018 July 6 as part of Observing Program 15285.  We used the H$\alpha$ image from 2018 rather than the Hubble Heritage H$\alpha$ image both because it gives a somewhat longer baseline for the proper motions and because, like the Hester 1997 image, it excludes the [N~II] lines.  The 2018 image will be discussed along with images in other lines in a subsequent paper.  The Hubble Heritage Project page http://heritage.stsci.edu/2015/29/index.html includes an overlay of the WFPC2 field on the Heritage mosaic, as well as an animation of the motions seen in H$\alpha$.   Fig.~\ref{pm_ha} shows the 1997 images in H$\alpha$ and [O~III] and indicates the positions where the proper motions were measured.  We used a {\sc python} package developed by K.~Grishin to correct the astrometry in the HST images by first detecting sources, then matching them against the Gaia DR2 catalog and re-computing the projection matrix using a nonlinear $\chi^2$ fitting, and finally updating World Coordinate System keywords in the FITS headers. The resulting astrometric accuracy was of an order of 0\farcs008--0\farcs015. Then we re-projected all updated HST images from different epochs using SWarp \citep{bertin02} to the same center position and resampled them to 0\farcs04 pixels.

The complexity of the images defeated our attempts at a fully automated 2D proper motion measurement, so we adopted a semi-automated procedure.  We identified sharply defined filaments that were relatively simple in the H$\alpha$ and [O~III] pairs of images and computed the cross-correlations for strips perpendicular to the filaments.  Oblique motions are also possible in principle, though we see no morphological evidence for motions at other angles.  We avoided filaments whose appearance changed drastically between the first and second exposure.

Fig.~\ref{pm_ha} displays the velocities inferred from the proper motions in H$\alpha$ and [O~III] assuming a distance of 735 pc \citep{fesen18}.  Given the time intervals between the image pairs, a single pixel shift corresponds to 7.93 \kms for [O~III] and 6.71 \kms for H$\alpha$.  In earlier studies the distance uncertainty was larger than the uncertainty in proper motion, but \citet{fesen18} give a distance uncertainty smaller than 4\%.  We did not attempt to measure the cross-correlations to sub-pixel accuracies because the 1997 H$\alpha$ image has already been resampled from 0\farcs1 to 0\farcs04.  Therefore, the measured values are multiples of the values for single pixel shifts. 

There are relatively few cases in which the same feature can be measured in both the [O~III] and H$\alpha$ images.  An especially striking case is the sharp H$\alpha$ filament near 20$^h$45$^m$44$^s$ and +30$^\circ$59$^\prime$30\arcsec, which is invisible in [O~III] and shows speeds above 300 \kms.  An even fainter filament is visible near the southern edge of the image at RA 20$^h$45$^m$39$^s$ with speeds around 200 \kms. It is not visible in [OIII], but seems to be an extension of the north-south [O~III] filament that stretches from +30$^\circ$58$^\prime$30\arcsec to 30$^\circ$59$^\prime$ and  whose speed increases from 118 \kms in the north to 174 \kms at its southern end.  These H$\alpha$ filaments are nonradiative shocks typical of those that produce the X-ray emission of the Cygnus Loop \citep{ghavamian01, salvesen09, medina14}, one of which was observed just to the north of our target region \citep{raymond80}.  They are projected onto the field we are observing but are not part of the radiative shock structure.  The fast [O~III] shock is a very incomplete shock, in which the gas has not yet cooled enough to produce strong H$\alpha$ emission \citep{raymond88}.

In the northern part of the WFPC2 images, corresponding to the region observed by Binospec, the crisp [O~III] filaments to
the west have speeds of 111--126 \kms, while the long, sharp [O~III] filament to the east is moving at 126--150 \kms.  Where H$\alpha$ proper motions can be measured at closely corresponding positions, they show similar speeds, but there is also considerable H$\alpha$ emission with much lower speeds, around 60-87 \kms.  We will take the typical [O~III] velocity of 130 \kms as the shock speed from now on, recognizing that there are variations within the region observed and that there is additional emission from slower shocks superposed, particularly toward the eastern end of the Binospec region.


\subsection{Average Spectrum}

Table 1 presents the unweighted average of the Binospec spectra, averaged over the 110 arcsec region where the emission is bright.  The lines were identified from the list of \citet{fesenhurford}.  Weak lines, of order 1-3\% of H$\beta$, have significant uncertainties due to the number of counts and determination of the continuum level, and they may be uncertain by as much as a factor of 2.



From the H$\alpha$/H$\beta$ ratio and the extinction curve of \citet{cardelli88} we estimate a reddening value of E(B-V)=0.20 or a little higher.  That is higher than the E(B-V)=0.08 typically adopted for the Cygnus Loop, but the radiative shocks along the western edge of the Cygnus Loop mark the encounter of the blast wave with a dense cloud that is apparent in decreased star counts and in H I 21 cm \citep{leahy02}.  The relatively high reddening is in line with that of the western cloud.  \citet{fesen18} used the interstellar dust map of \citet{green15} to show that the reddening in the direction of the western cloud increases gradually to a value near 0.1 at about 700 pc, then jumps to 0.55 by 1 kpc.  We use E(B-V)=0.20 to deredden the fluxes in Table 1 and Figures~\ref{o3ha} and \ref{dene_los}.  The H$\alpha$ to H$\beta$ ratio varies along the slit, suggesting a variation in E(B-V) of order 0.05.

\subsection{Models}

Model spectra were computed with an updated version of the code of \citet{raymond79} and \citet{coxraymond}.   In particular, we have added [Fe VI] and [Fe VII] lines using atomic rates from CHIANTI \citep{dere97, delzanna15}.  The code assumes a 1D steady flow, using the Rankine-Hugoniot jump conditions to find the postshock gas parameters.  Then it uses the fluid equations to compute the density, temperature, and velocity as the gas cools. The perpendicular component of the magnetic field is assumed to be frozen in, and it is compressed with the gas as it cools.  Time-dependent ionization calculations including photoionization are used to compute the cooling rate and the emissivities of spectral lines.

Two models for 130 km/s shocks are shown in Table 1.  The speed was chosen to match the average proper motion, and the preshock density and perpendicular magnetic field were set to 6 $\rm cm^{-3}$ and 4 $\mu$G to match the ram pressure and postshock density derived below. The preshock gas was taken to be ionized, with a temperature of $10^4$ K.  The elemental abundances were taken to be H:He:C:N:O:Ne:Mg:Si:S:Ar:Ca:Fe = \\ 12.0:10.93:8.52:7.96:8.82:7.92:7.42:7.52:7.20:6.9:6.3:7.5.  The electron and ion temperatures were taken to be equal, in accord with the indications of full thermal equilibration in the Cygnus Loop \citep{medina14, raymond15}. 

\begin{figure*}
  \center
\includegraphics[width=\hsize,trim={0 0 0 18cm},clip]{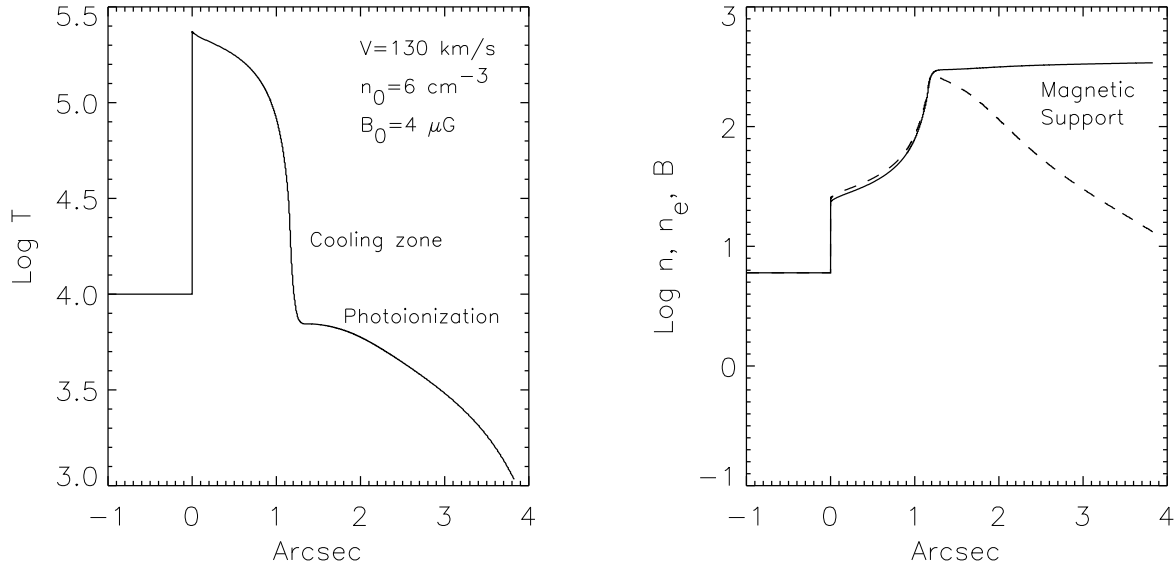}
\caption{Model of the temperature and density structure of a 1D, steady-flow shock.  The left panel shows Log T as a function of
distance from the shock assuming 735 pc for the distance to the Cygnus Loop.  In the right panel the solid line shows the hydrogen density in $\rm cm^{-3}$ (solid line) and the electron density (dashed).  The magnetic field strength in $\mu$G is 2/3 the density.
\label{model_130}
}
\end{figure*}

Figure~\ref{model_130} shows the computed temperature and density structure.  The [O~III] emission comes mostly from the region where the temperature drops steeply from $10^5$ K to about $2 \times 10^4$ K, while the H$\alpha$ and [S~II] are produced in the photoionized region below $10^4$ K.  That zone is a few times $10^{16}$ cm thick, which corresponds to a few arcseconds at the distance of the Cygnus Loop. The right panel shows the increase in density that maintains pressure equilibrium as the temperature drops.  After the gas reaches about $10^4$ K, magnetic pressure dominates, and the density remains constant as the gas continues to cool and recombine.

Two models are given.  Model 130 is a ``complete'' shock, in which the gas was allowed to cool to 1000 K.  In Model 130i the integration was terminated at 6700~K, corresponding to a shock that had just encountered the dense cloud 1000 years ago.  This is a crude, but convenient, single-parameter way to account for the finite age of a shock, and it reduces the emission in the coolest lines -- the Balmer lines, [O~I] and [N~I].  Such incomplete shocks are seen as filaments with anomalously large [O~III]/H$\beta$ ratios in the Cygnus Loop and other SNRs \citep{raymond88}, and they are apparent in Figure 1 as regions bright in [O~III] with little corresponding H$\alpha$ or [S~II]. 

The model H$\beta$ intensities pertain to a planar shock viewed face-on.  They are several times smaller than the observed value, indicating modest limb brightening. However, the average intensity includes mostly regions of relatively faint emission, and brightest filaments are considerably brighter, indicating more extreme limb-brightening at tangencies to the line of sight (LOS) \citep{hester87}.

We conclude that the 130 \kms\  models provide as good a match to the data as could be hoped for, given the uncertainties in atomic rates and elemental abundances and measurement errors in the fainter lines.  That is in spite of the fact that the real shock morphology is far more complex than the assumed 1D steady flow and the fact that such potentially important processes as dust cooling, dust destruction, thermal conduction, and cosmic-ray acceleration are not included.  The reasonable agreement between the 1D model and the observed spectrum for the higher ionization states may reflect the fact that there are several free parameters, though we have chosen values consistent with the proper motion and the density and magnetic field derived by other means below.  It may also indicate that the basic physics of the models is conservation of energy in the cooling gas, and to first order that determines the average emission spectrum, though complications such as thermal instabilties significantly modify the flow and emission at any given point or time \citep{innes92}.  The underprediction of the low-temperature lines of [O~I] may indicate that slower shocks seen in H$\alpha$ but not [O~III] contribute significantly, or that the complex structure seen in H$\alpha$ leads to a lower ionization state than predicted by the models.  The underprediction of [O~III] favors the ``incomplete'' model, which also gives better agreement with the [Ne~IV] and [Ar~IV] lines.  A slightly higher shock speed, $\sim$ 140 \kms, would also improve the agreement for those lines, but it would make the underprediction of the [O~III] lines worse.

The 130 \kms incomplete shock model is not a unique interpretation of the data.  The proper motions indicate that a range of shock speeds is present.  Figure~\ref{pm_ha} shows nonradiative shocks faster than 300 \kms, but they are not apparent in the region observed by Binospec, except for the presence of a faint [Fe X] line.  Smaller proper motions are also apparent in H$\alpha$, but we do not know whether they are actual shock speeds or the result of the turbulent motions suggested by the complex morphology of the H$\alpha$ emission.  Similarly, we have chosen to truncate the incomplete model at 6700 K because a cutoff a few hundred degrees lower gives only a small increase in the [O III] to H$\beta$ ratio, while a cutoff a few hundred degrees higher drastically reduces the emission in the coolest.  Given the limitations to the model discussed above, we have presented a model with the smallest number of free parameters that gives a reasonable match to the observations, because one of our goals is to determine whether the simple 1D can be used to determine shock speeds and abundances in shocks that are not spatially resolved.

\subsection{Shock Age}

The largest discrepancy between the M130 model and the observed spectrum, apart from the underprediction of the cool [O~I] lines, is the underprediction of the [O~III]/H$\beta$ ratio.  Model M130i is truncated at about 1000 years, and it increases the [O III]/H$\beta$ ratio by 50\%, bringing it into reasonable agreement with the observed value.  It also improves the agreement with the [Ne~IV] and [Ar~IV] lines, but leads to more severe overprediction of the [Fe~VI] and [Fe~VII] lines relative to H$\beta$.  This discrepancy could be overcome if Fe is depleted onto dust grains, as discussed in the next section.

The better agreement of the model truncated at 6700~K suggests that the shock only reached the dense cloud relatively recently in this location, roughly 1000 years ago, though the agreement is still not perfect.  If the shock encountered a gradually rising density rather than a density jump at the cloud, the longer cooling times at lower densities would mean a larger age.  The age could also be estimated from the degree of indentation of the shell by comparing the shape with a sphere.  Within very large uncertainties, the indentation seems to be 10\% to 20\% of the radius, which for a remnant age of 21,000 years \citep{fesen18} indicates a shock-cloud interaction age of order 2,000 yr.  However, the apparent indentation is likely to be exaggerated by projection effects.
  
An alternative explanation for the excess [O~III]/H$\beta$ might be related to the complex structure seen in the H$\alpha$ image in Figure~\ref{pm_ha}.  If the recombining gas is concentrated into dense clumps with a modest filling factor, some of the ionizing photons could escape into the interior of the Cygnus Loop.  That would reduce the number of recombinations per H atom, and therefore reduce the H$\beta$ emission.  It is not obvious how the intensities of the [N~I] and [O~I] lines would be affected, because the ionization state would be lower, but the heating that balances their emission would also be reduced.  

\subsection{Dust Destruction}
  
An interesting feature of Table~\ref{spectrum} is that [Fe~X] is observed, while the models predict that it is undetectable.  The [Fe X] must be formed in the hotter X-ray emitting gas that envelops the optical filaments. It is produced by nonradiative shocks, in particular the shocks faster than 300 \kms seen in Figure~\ref{pm_ha} that produce the X-ray emission observed in the region \citep{levenson02}.

On the other hand, the [Fe VI] lines are overpredicted relative to H$\beta$ by the models, while one of the [Fe VII] is overpredicted and the other matches the observed value.  These faint lines are subject to uncertainties of order a factor of 2, but all are clearly detected.  The simplest interpretation is that Fe is heavily depleted onto grains in the preshock gas, and that sputtering returns about 1/3 of it to the gas phase before the gas cools below about $2 \times 10^5$ K.  If the shock speed is significantly faster than 130 \kms, the [Fe~VI] and [Fe~VII] lines would be predicted to be stronger, and a correspondingly more severe depletion would be required.

The indication that 1/3 of the Fe is liberated from grains in the high-temperature gas just behind the shock contrasts with the conclusions of \citet{dopita16}.  They found from spectra of the Large Magellanic CLoud (LMC) supernova remnant N49 that only 10\% of the Fe had been liberated from grains at the temperatures where [Fe~VI] and [Fe~VII] are formed, while about 70\% of the Fe is returned to the gas phase by the time the gas has cooled to the temperatures where [Fe~II] is formed.  The [Fe~VI] and [Fe~VII] line ratios to H$\beta$ are similar in the N49 spectrum of Dopita et al. and our Cygnus Loop spectrum.  \citet{dopita16} modeled N49 with a shock speed of 250 \kms and LMC abundances, which predicted [Fe~VI] and [Fe~VII] lines more than 5 times stronger than their 130 \kms\/ models.  Therefore, the lower Cygnus Loop shock speed determined from proper motions accounts for some of the difference in interpretation.  \citet{sankrit14} found iron to be depleted by about a factor of 2 from the IR lines of [Fe~II] and [Fe~III] in the eastern Cygnus Loop.  Similar depletions of iron at the temperatures of [Fe~VI], [Fe~VII] and [Fe~II], [Fe~III] would suggest that rapid sputtering in the immediate postshock region accounts for a major part of the grain destruction.  Models indicate that grain-grain collisions at lower temperatures are also important \citep{slavin15}, and inertial sputtering may return some of the shattered grain material to the gas phase.  

It is also possible that there are faster shocks along the line of sight that produce [Fe~VI] and [Fe~VII], but that have not yet become radiative and therefore do not appear in the [O~III] and H$\alpha$ images used to measure the proper motions.  Such shocks would be analogous to the nonradiative shocks that produce the [Fe~X] line, but somewhat slower, as observed in the northeastern Cygnus Loop \citep{blair05}.  They would have to occupy a narrow range in shock speed and age because of the rapid cooling of gas near $2 \times 10^5$ K.  That hypothesis could be tested by searching for higher-temperature lines such as [Ne V], but the reddening would make it difficult to observe UV lines such as O~VI.

\begin{figure}
\includegraphics[clip,trim={0.55cm 0 0.2cm 0},height=0.8\vsize]{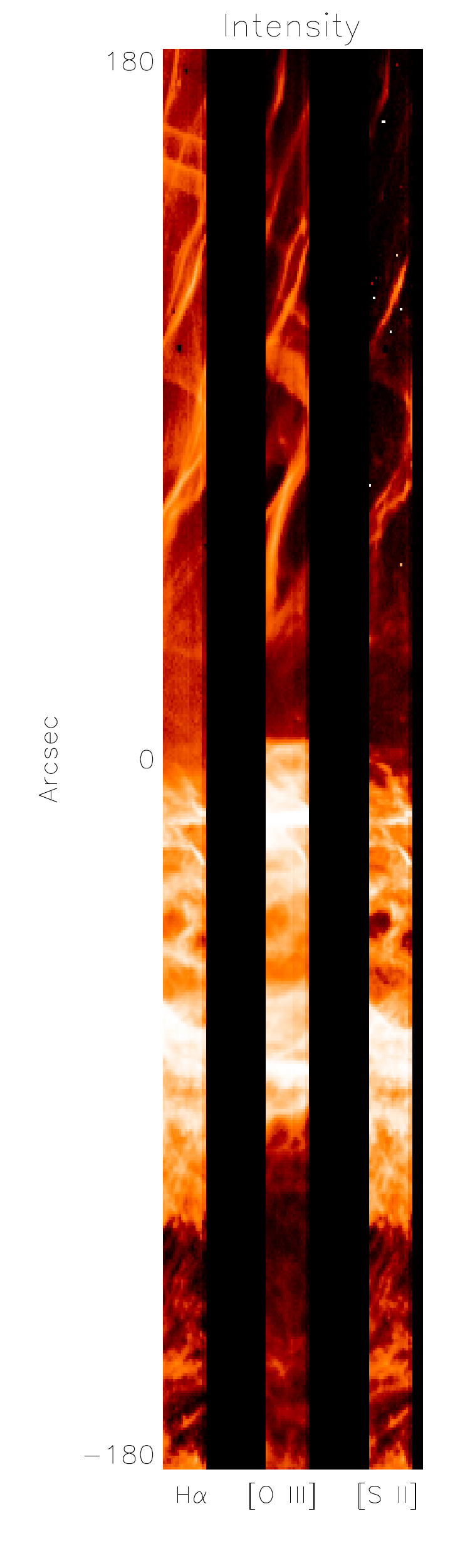}
\includegraphics[clip,trim={2.55cm 0 0.8cm 0},height=0.8\vsize]{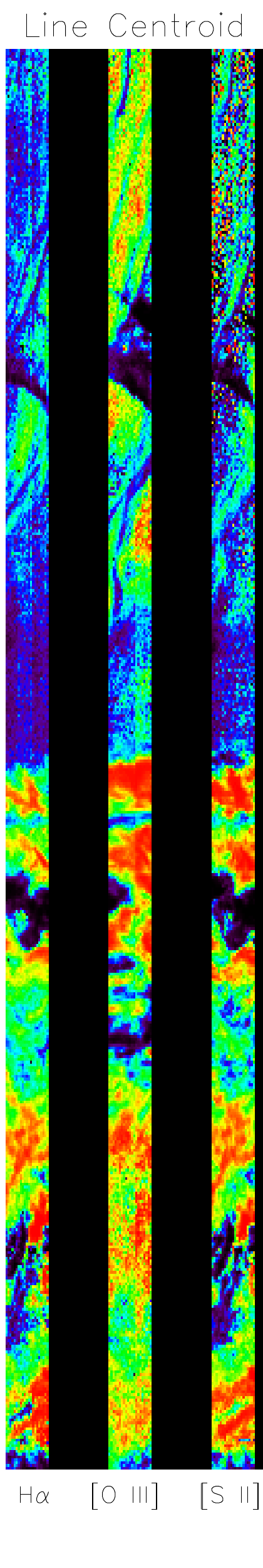}
\caption{2D representations of the intensities and velocity centroids of the H$\alpha$, [O III] and [S II] lines from the Binospec spectra. The vertical axis is the spatial dimension along the slit and the horizontal axis corresponds to the 23 individual spectra with the offsets scaled to match the scale of the vertical axis.  The 0\arcsec position corresponds to the bright [O III] filament near the center of the image in Figure~\ref{overlay}, and the bright [O III] filament on the eastern side of the image is at about -65\arcsec .  The intensity range is 0 to 
{\bf $2 \times 10^{-14}~\rm erg~cm^{-2}~s^{-1}~arcsec^{-2} $} and the velocity scale ranges from -50 to +50 \kms.
\label{2D_IV}
}
\end{figure}

\subsection{Detailed Structure}

We turn now to the detailed structure along the Binospec slit and across the neighboring slits.  Figure~\ref{2D_IV} shows false color representations of the intensities and velocity centroids of the H$\alpha$, [O~III] and [S~II] lines.  They are obtained from Gaussian fits to the line profiles at each spatial pixel in each spectrum.  It is apparent that the H$\alpha$ and [S~II] structures are very similar to each other, as expected from the 1D models.  The [O~III] intensity structure is considerably different, but its velocity range is similar to those of the cooler lines, with a spread of about $\pm$20 \kms.

\begin{figure*}
  \center
    \includegraphics[width=\hsize]{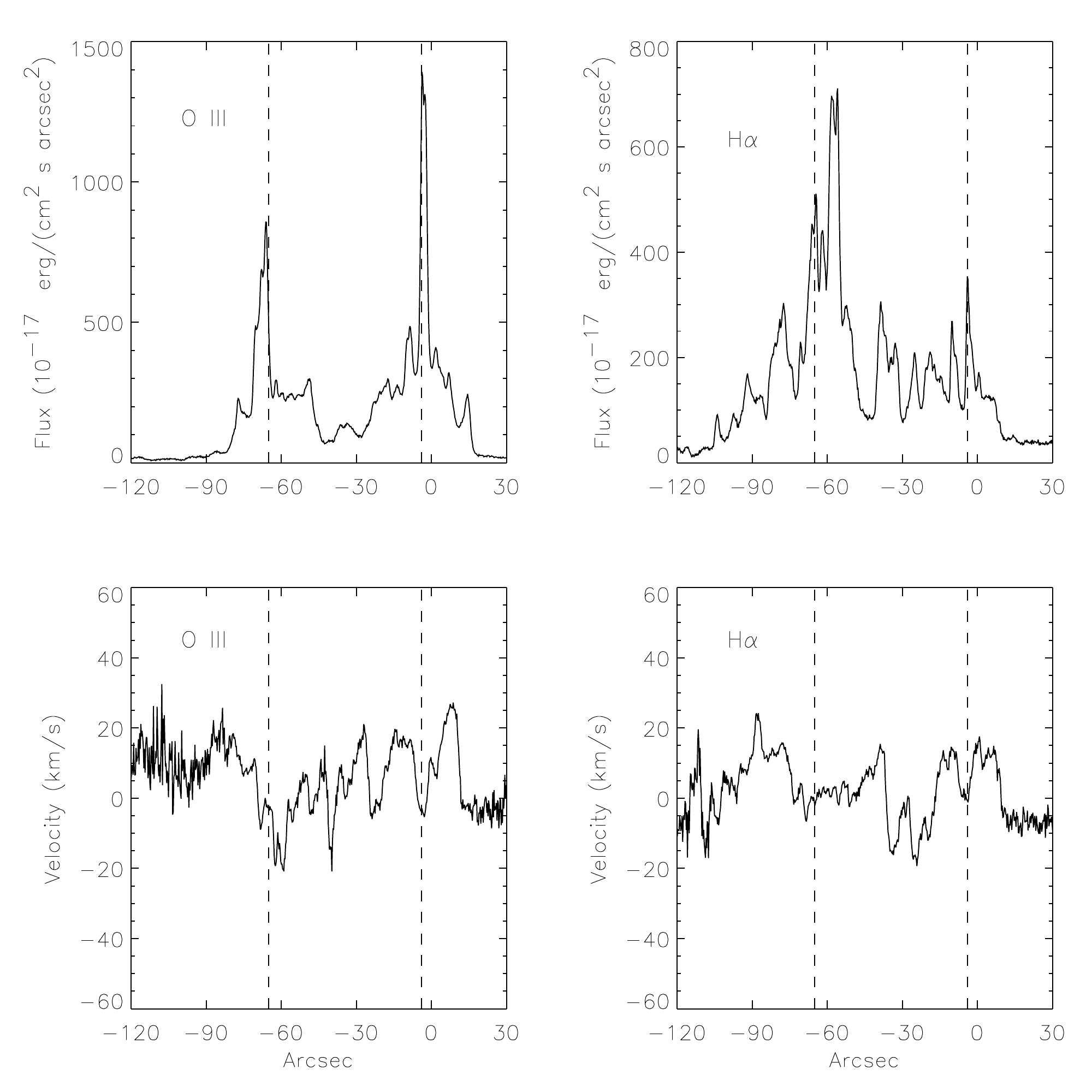}
\caption{Intensities and velocity centroids of the [O III] and H$\alpha$ lines along a single slit near the center of the observed band.  Positions are shown in arcseconds.  The 0\arcsec position corresponds to the bright [O III] filament near the center of the image in Figure~\ref{overlay}, and the bright [O III] filament on the eastern side of the image is at about -65\arcsec .  The fluxes have not been corrected for reddening.  There is little correlation between the intensities or velocities of the two lines. The dashed lines indicate the positions of the strongest [O III] peaks in the four plots.
\label{o3ha}
}
\end{figure*}

To show the structure more quantitatively, we present plots of observed and derived quantities along the central slit position.  Figure~\ref{o3ha} shows the intensities of the [O~III] and H$\alpha$ lines as functions of position along the slit.  The position increases toward the west, and zero is at 20$^h$ 45$^m$ 37.544$^s$, +31$^\circ$ 0$^\prime$ 9\farcs4.  The [O~III] and H$\alpha$ lines are bright within the same region along the slit, but there is remarkably little similarity in the intensity structure of the two lines.  As expected from Fig.~\ref{pm_ha} (bottom), there are two well-defined peaks where the slit crosses the bright [O~III] filaments, while the H$\alpha$ intensity is uncorrelated with the [O~III] except that both occupy more or less the same section of the slit.

Figure~\ref{o3ha} shows the observed velocities, with zero corresponding to about $V_{LSR}$ = 12 \kms , close to the velocity of the western cloud obtained from 21 cm observations \citep{leahy02}.  The velocity centroids are close to zero at the [O~III] peaks, as expected if the peaks are tangencies of a rippled sheet to the line of sight \citep{hester87}.  Not all of the zero-velocity points correspond to intensity peaks, perhaps because positive and negative velocity contributions happen to cancel out at those positions.  H$\alpha$ does not show an obvious relationship between intensity peaks and velocity zeros, suggesting that the H$\alpha$ strucure has more to do with intrinsically bright clumps or filaments than with tangencies of a smooth sheet to the line of sight.

\subsubsection{Density and LOS depth}

\begin{figure}
  \center
    \includegraphics[width=\hsize]{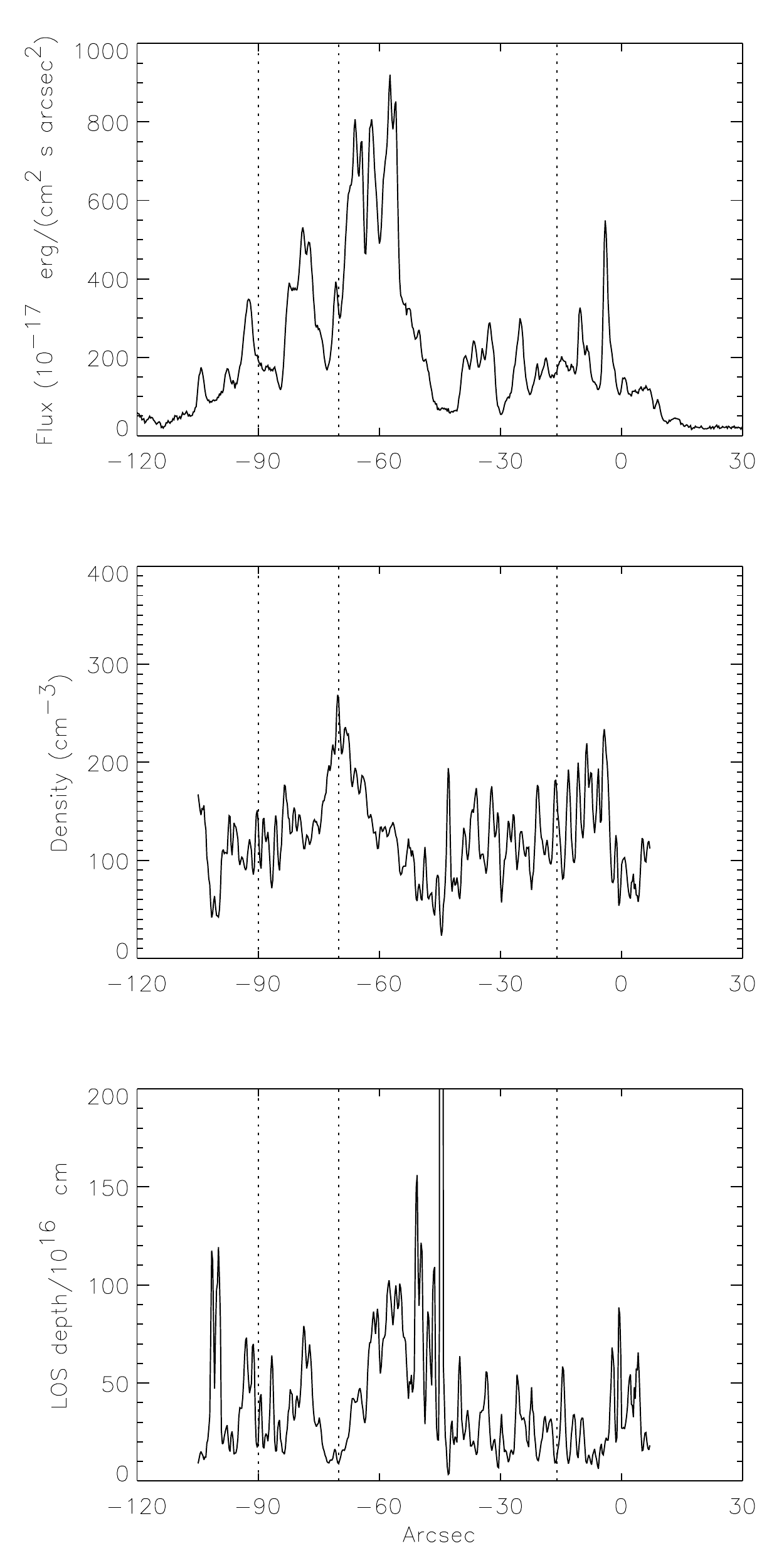}
\caption{[S II~] fluxes and electron densities derived from the [S II] doublet ratio, along with line-of-sight depths determined from the intensity and density.  Fluxes have not been corrected for reddening.  The temperature is taken to be 7000 K and sulfur is taken to be entirely singly ionized (S II) in the region where it is emitted.  Dashed lines guide the eye to show alignment or lack thereof among intensity, density, and LOS depth.
\label{dene_los}
}
\end{figure}

Figure~\ref{dene_los} shows the flux in the [S~II] doublet along the same slit, along with the electron density derived from the intensity ratio.  We assume a temperature of 7000 K.  The bottom panel shows the depth along the line of sight obtained from the intensity and the density.  For that estimate we assume that all the sulfur is singly ionized (S~II), so the actual LOS depth  will be somewhat larger.  LOS depths of {\bf $10^{18}~\rm cm^{-3}$} are consistent with the limb brightening inferred above by comparing the H$\beta$ surface brightness with models.

\subsubsection{Temperature}

\begin{figure}
  \center
    \includegraphics[width=\hsize]{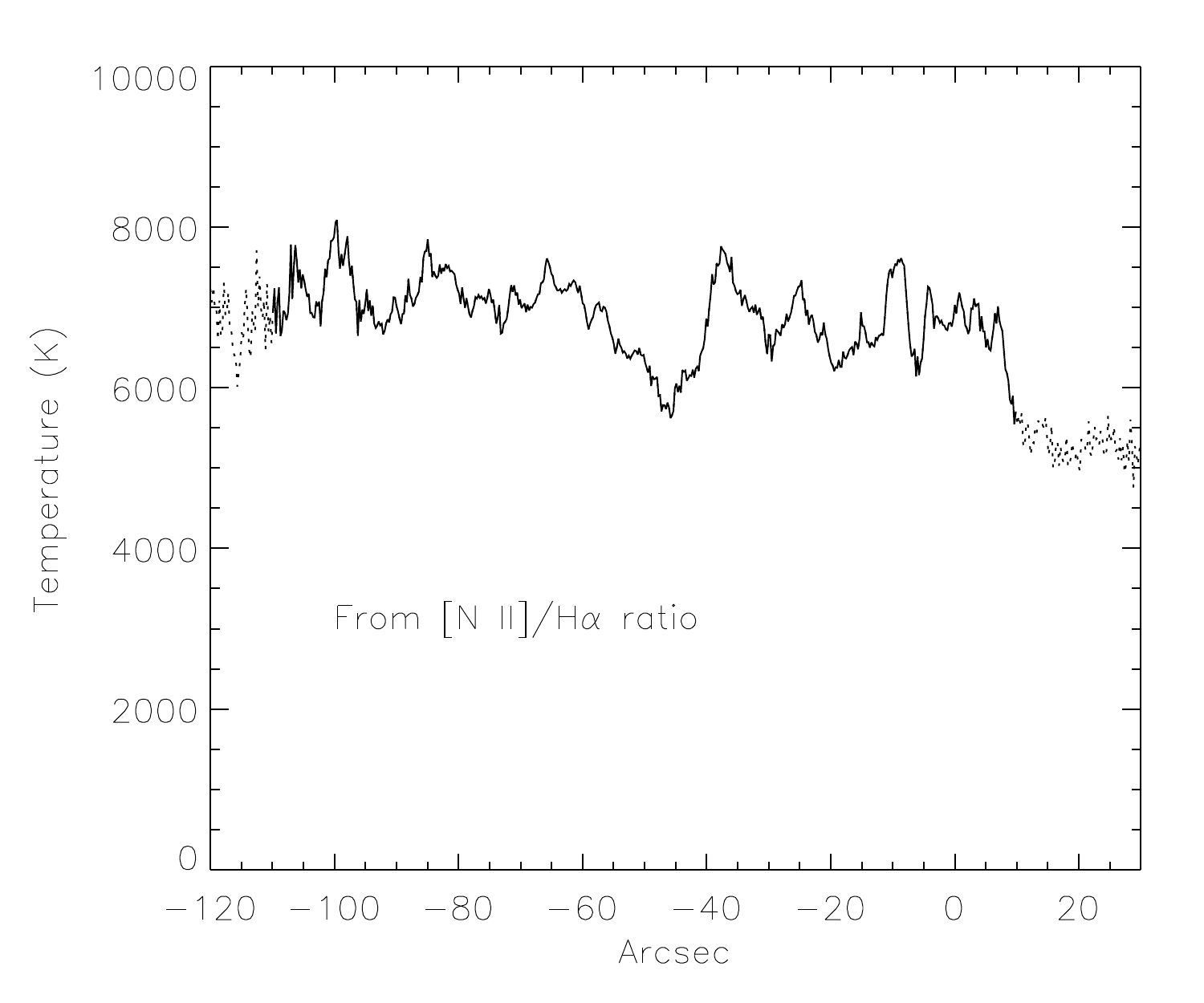}
\caption{Temperatures inferred from the ratio of [N II] intensity to H$\alpha$.  They assume that all the nitrogen is singly ionized in the region where the H$\alpha$ is produced, so they slightly underestimate the temperature.  The curve is dotted where the emission is faint and the uncertainties are higher.
\label{t_n2}
}
\end{figure}

Figure~\ref{t_n2} shows a temperature estimate obtained from the ratio of the [N~II] lines to H$\alpha$.  The H$\alpha$ is produced by recombination while [N~II] is produced by collisional excitation, so the ratio is proportional to the Boltzmann factor in the excitation rate, $e^{-22,000/T}$.  Since nitrogen is fairly tightly coupled to hydrogen by charge transfer \citep{butlerdalgarno}, it is reasonable to assume that nitrogen is at least singly ionized in the region where hydrogen recombination occurs.  Nitrogen is more highly ionized above about 25,000 to 30,000 K, but relatively little emission in either line comes from those regions, so we expect a modest underestimate of the temperature in the H$^+$ - N$^+$ zone.  We assume a nitrogen abundance of 7.96 as in the shock models and use atomic rates from CHIANTI \citep{dere97, delzanna15}. The derived temperature range of 6000 to 8000 K is in good agreement with model predictions, but it should be remembered that the uncertainty in N abundance carries over into the temperature estimate.   

As a check, we can compare the [N~II] line ratio 5755/6584 from Table 1 with predictions from CHIANTI.  The observed value of 0.0165 corresponds to 10,500 K, and an uncertainty of 20\% due to statistics, background subtraction, and extinction would imply a range 10,000-11,000 K. As mentioned above, the [N II]/H$\alpha$ ratio should give a modest underestimate of the temperature, while the 5755/6584 gives an average biased toward higher temperatures by the steep increase of the $\lambda$5755 emissivity due to the $e^{-47,000/T}$ Boltzmann factor. Thus the difference between the two estimates of temperature is partly due to the contribution to H$\alpha$ from regions where N is more than singly ionized, and partly due to the fact that the two temperature estimates are differently weighted averages.  

\subsubsection{Ram pressure}
\begin{figure}
  \center
\includegraphics[width=1.0\hsize]{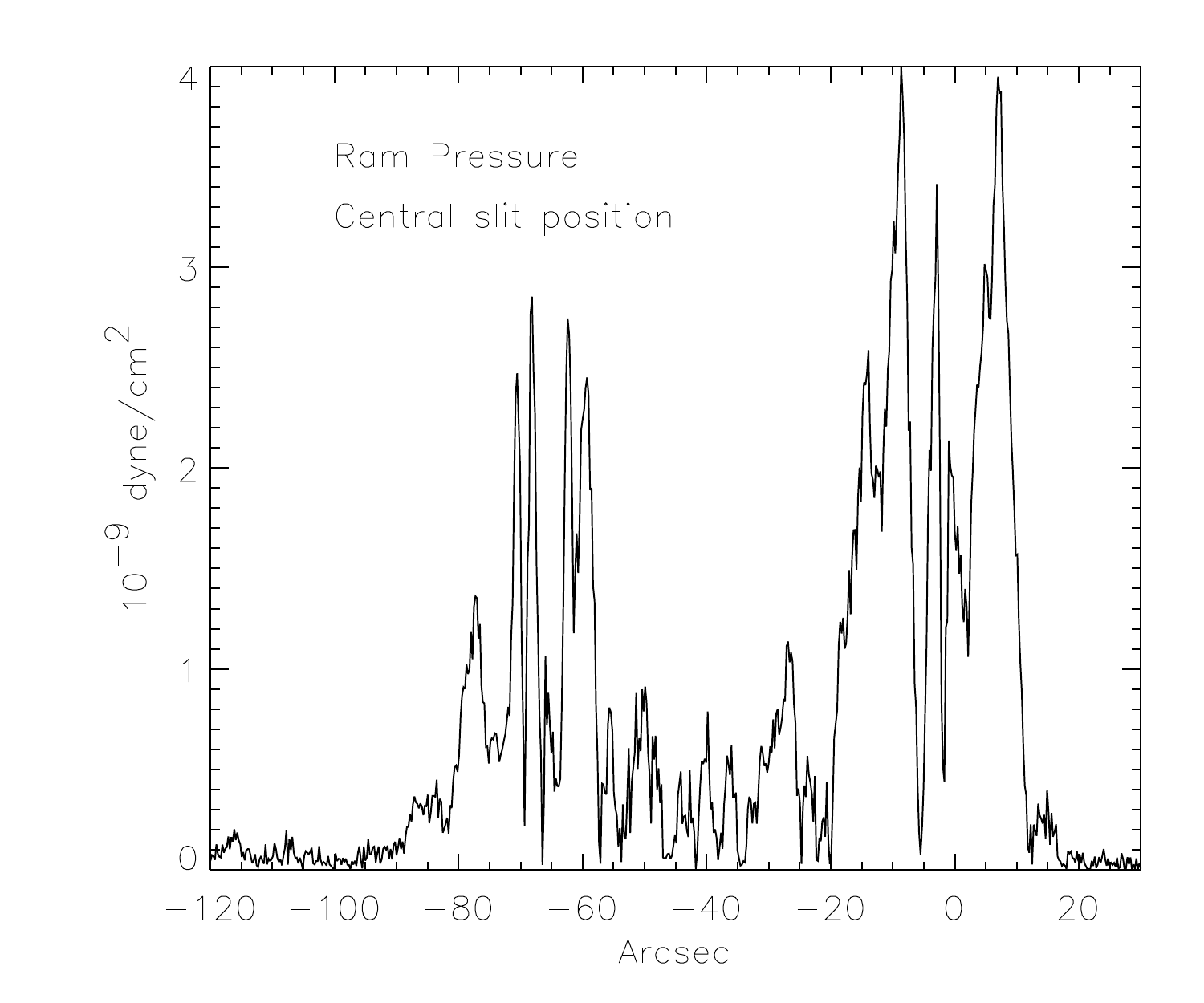}
\caption{Shock ram pressure derived from the [O III] intensity and line centroid as described in the text.  These values are lower limits, but they are expected to be close to the actual values at the highest peaks.
\label{rampress}
}
\end{figure}

The ram pressure, $\rho_0 V_S^2$, is a fundamental parameter of the shock.  Since it approximately equals the driving pressure, it tends to be constant over large regions, implying that the shock speed varies as $n_0^{-1/2}$, and it relates the parameters of the radiative and nonradiative shocks.  Figure~\ref{rampress} shows the shock ram pressure as a function of position along the slit.  It is obtained by the method proposed in \citet{raymond88}.  According to the 1D models, shocks in the 100--150 \kms\/ range produce about 0.55$\pm$0.05 [O III] photons per H atom that passes through the shock, so the surface brightness is proportional to $n_0 V_S$.  If the shock is viewed at an angle to the line of sight, the brightness, $I_{OBS}$ increases by a factor of 1/cos$\theta$, where $\theta$ is the angle between the line of sight and the shock normal. The emission will be Doppler shifted by $V_{DOPP} = V_S $cos$\theta$, so cos$\theta$ cancels out in the product $I_{OBS} V_{DOPP}$ of observed intensity times Doppler shift.  Therefore, $I_{OBS} V_{DOPP}$ is simply a constant times $n_0 V_S^2$, or the ram pressure $\rho_0 V_S^2$.  We take the number of [O III] photons per H atom to be 0.55 for Figure~\ref{rampress}.

This procedure will obviously produce underestimates of the ram pressure a) if $\theta$=90$^\circ$ so that the Doppler shift is zero, or b) if the line of sight passes through approaching and receding shocks whose Doppler shifts partly cancel when the centroid of an unresolved line is measured.  Examples of a) are clearly seen where the Doppler velocity crosses between positive and negative in Figure~\ref{o3ha}, and the complex rippled sheet structure suggested by Figure~\ref{overlay} implies that b) must also be important in many places.  However, at any position where either positive or negative velocity dominates, $I_{OBS} V_{DOPP}$ should give a reliable estimate of the ram pressure.  Such positions should occur where the LOS does not pass through the shock multiple times, so the highest apparent ram pressures in Figure~\ref{rampress} ought to be close to the true ram pressure.  We also note that the H$\alpha$ line width is 20-30 \kms after subtraction of the 82 \kms instrumental width measured from night sky lines, and the H$\alpha$ width includes a thermal width of order 20 \kms.  Hence, Doppler velocities substantially larger than the $\approx$ 20-25 \kms seen in Figure~\ref{o3ha} are rare.  We therefore take $4 \times 10^{-9} \rm ~dyne~cm^{-2}$ to be a reasonable value.  It is about twice the value of $1.5 \times 10^{-9} \rm ~dyne~cm^{-2}$ in a partially radiative shock in the eastern Cygnus Loop obtained by \citet{raymond88} and comparable to the ram pressure implied by the parameters of the shock in the northeast \citep{hester94}.  It is about four times as high as the ram pressure in the blast wave from the parameters of \citep{fesen18}, which is expected when a blast wave encounters a dense obstacle \citep{hester94}.  High-resolution observations in the future could improve this estimate.

\subsubsection{Magnetic pressure}
\begin{figure}
  \center
    \includegraphics[width=1.00\hsize]{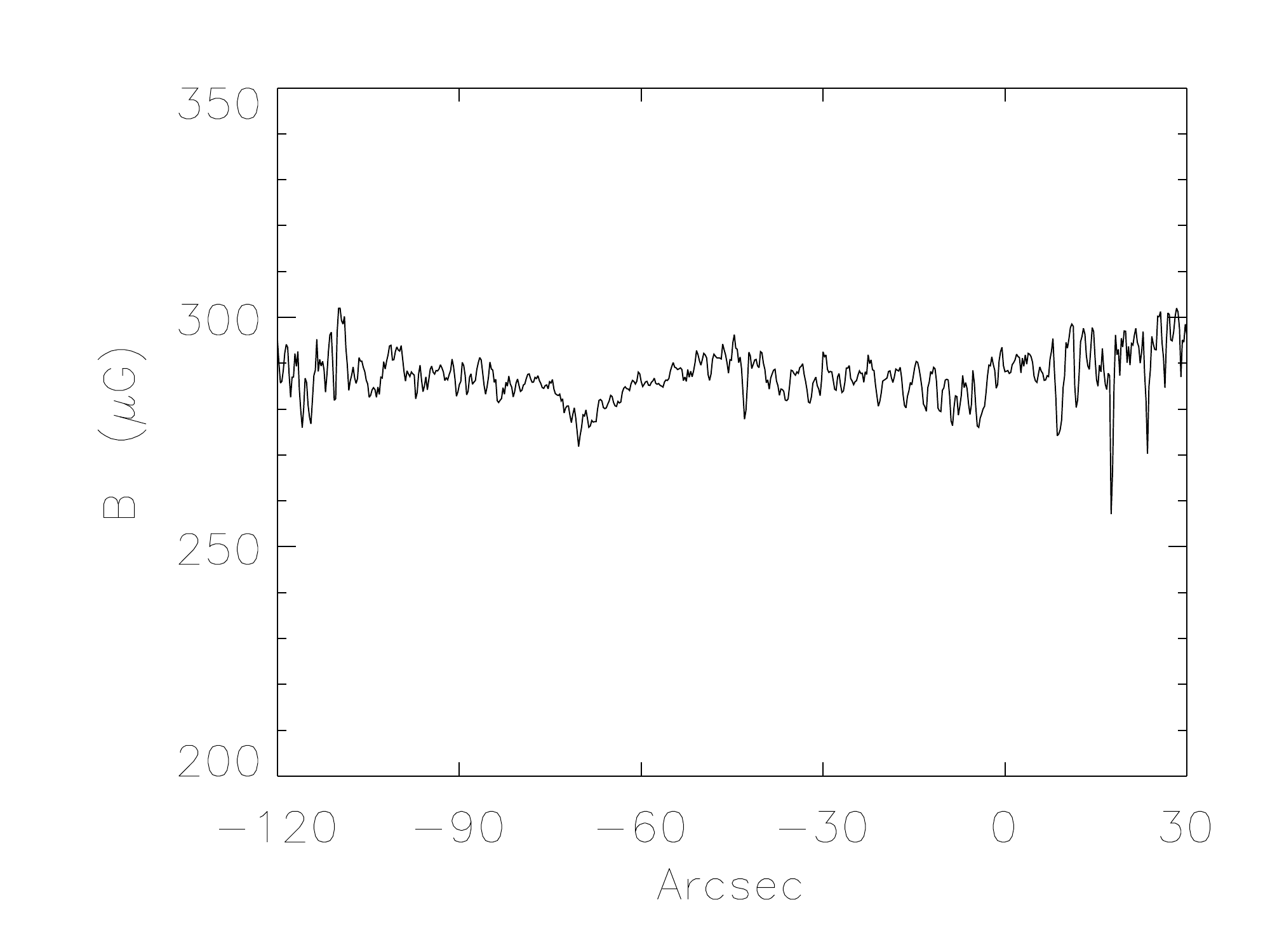}
\caption{Magnetic field in the region where H$\alpha$, [N II] and [S II] are produced, where T = 6000 to 8000 K and the electron densities are 100 to 250 $\rm cm^{-3}$.  These values were obtained by subtracting the thermal pressure from the ram pressure based on $P_B = B^2/8\pi$. 
\label{B}
}
\end{figure}

The flow behind the shock must be close to pressure equilibrium, since it is postshock pressure that drives the shock.  Thus the total pressure in the postshock gas should equal the ram pressure.  The thermal pressure is given by ($n_e+n_i)kT$ plus a moderate contribution from neutrals based on the ionization state from the 1D shock models.  We subtract the thermal pressure based on the densities and temperatures from Figures~\ref{dene_los} and ~\ref{t_n2} from the ram pressure to obtain the nonthermal pressure from magnetic fields and cosmic-rays, $P_B$ and $P_{CR}$.  The [S~II] lines are produced at a lower temperature than the [N~II] lines, but as discussed in connection with Figure~\ref{t_n2}, that figure gives a modest underestimate of the temperature, so it should give a reasonable estimate of the temperature in the [S~II] region.  The gas is strongly compressed as it cools from $\approx 2 \times 10^5$K to $10^4$K and the magnetic pressure scales as compression squared while adiabatic compression only increases the relativistic particle pressure as the 4/3 power.  Therefore, we expect that magnetic pressure dominates in the zone of singly ionized plasma.   Thus, we infer the magnetic field as a function of position shown in Figure~\ref{B}.

\subsubsection{Vorticity}

We next turn to an estimate of the vorticity of the flow, $\nabla \times V$, which is generated by interaction of the shock with density inhomogeneities.  This is important because it can wind up and amplify the magnetic field \citep{giacalone07, xulazarian, guo12} and because it can cascade to smaller-scale turbulence, accelerating particles and heating the gas \citep{zank15}.  Moreover, it could play an important role in explaining the glaring difference between the [O~III] and H$\alpha$ morphologies.  Measurements of the Doppler velocity over the two-dimensional regions shown in Figure~\ref{2D_IV} provide two of the six velocity gradients that enter the vorticity, and we use symmetry arguments to estimate the others.

\begin{figure}
  \center
    \includegraphics[trim={0.5cm 0 2.0cm 0},clip,width=1.07\hsize]{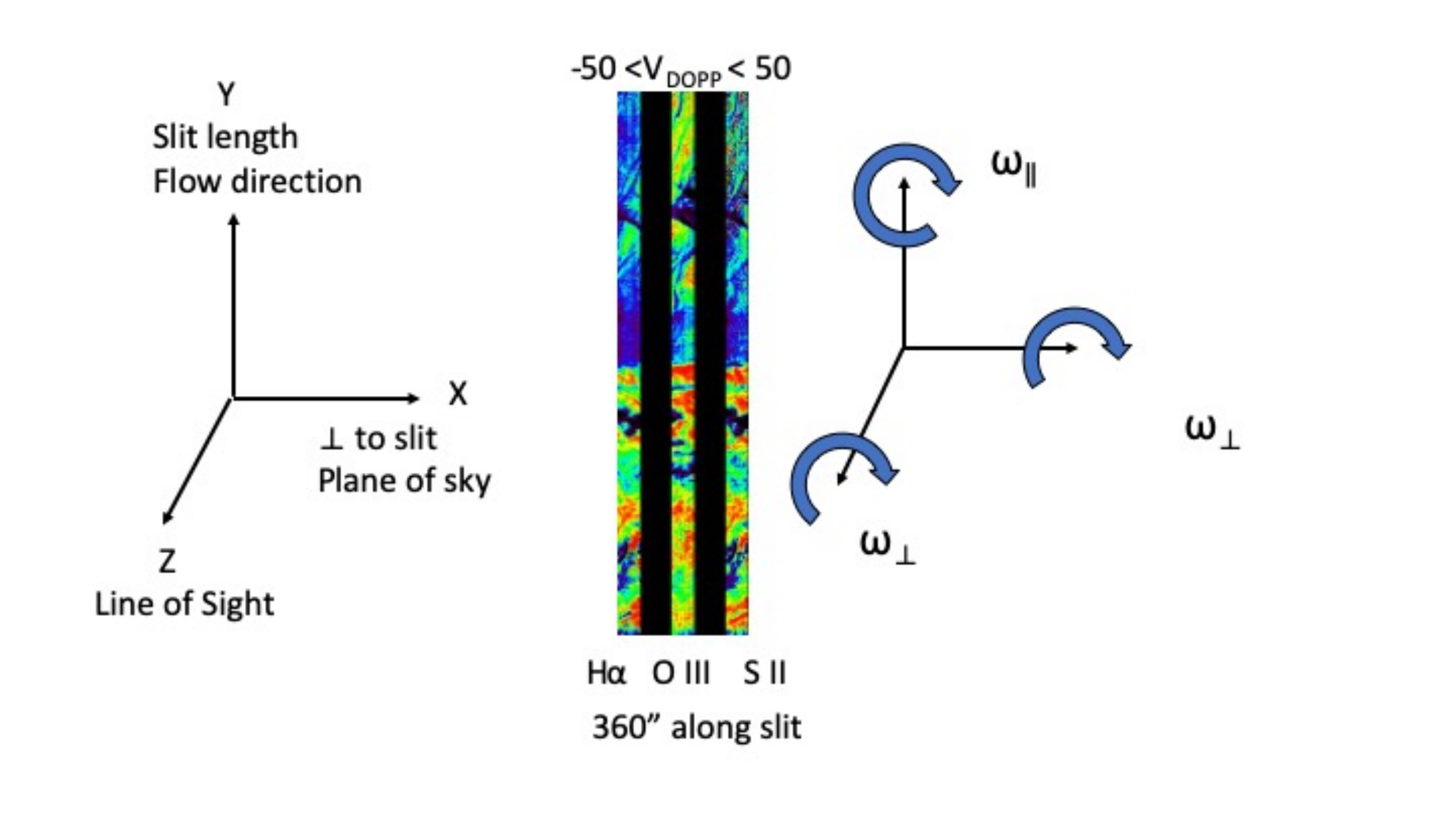}
\caption{Coordinate system showing the components of vorticity.
\label{schematic}
}
\end{figure}

We define the $x$, $y$ and $z$ directions to be perpendicular to the slit, along the slit, and along the line of sight (toward Earth), as shown in Figure~\ref{schematic}.  We measure $dV_z /dx$ and $dV_z /dy$ directly from the Doppler shift data.  The only preferred direction is the flow direction, along the $y$-axis, so the the $x$ and $z$ axes are equivalent.  That is, if we viewed the Cygnus Loop from Galactic south rather than from Earth, this region should look about the same.  Therefore, $dV_x /dz=dV_z /dx$ and $dV_x/dy=dV_z /dy$ in magnitude, though they are uncorrelated.  We therefore take the component of vorticity about the $y$-direction (shock normal or flow direction) to be $\sqrt 2 ~dV_z/dx$.  

The $y$-direction is preferred, so we cannot make the same symmetry argument for the terms involving $V_y$.  However, in a flow with deviations at a small angle $\alpha$ from the mean flow direction (as indicated by the gentle ripples in the [O~III] image), the velocity variations along and perpendicular to the flow are V~(1-sin$\alpha$) and V~cos$\alpha$, which are similar for small $\alpha$.  Therefore, we assume $dV_y /dz$ to be of the same magnitude as $dV_z /dy$, but uncorrelated, so that the component of vorticity perpendicular to the flow direction is  $\sqrt 2 ~dV_z/dy$.  This is admittedly a less robust estimate than that of the $y$-component of vorticity.

\begin{figure*}
  \centering
    \includegraphics[clip,trim={1.0cm 0 0.5cm 0},width=0.8\textwidth]{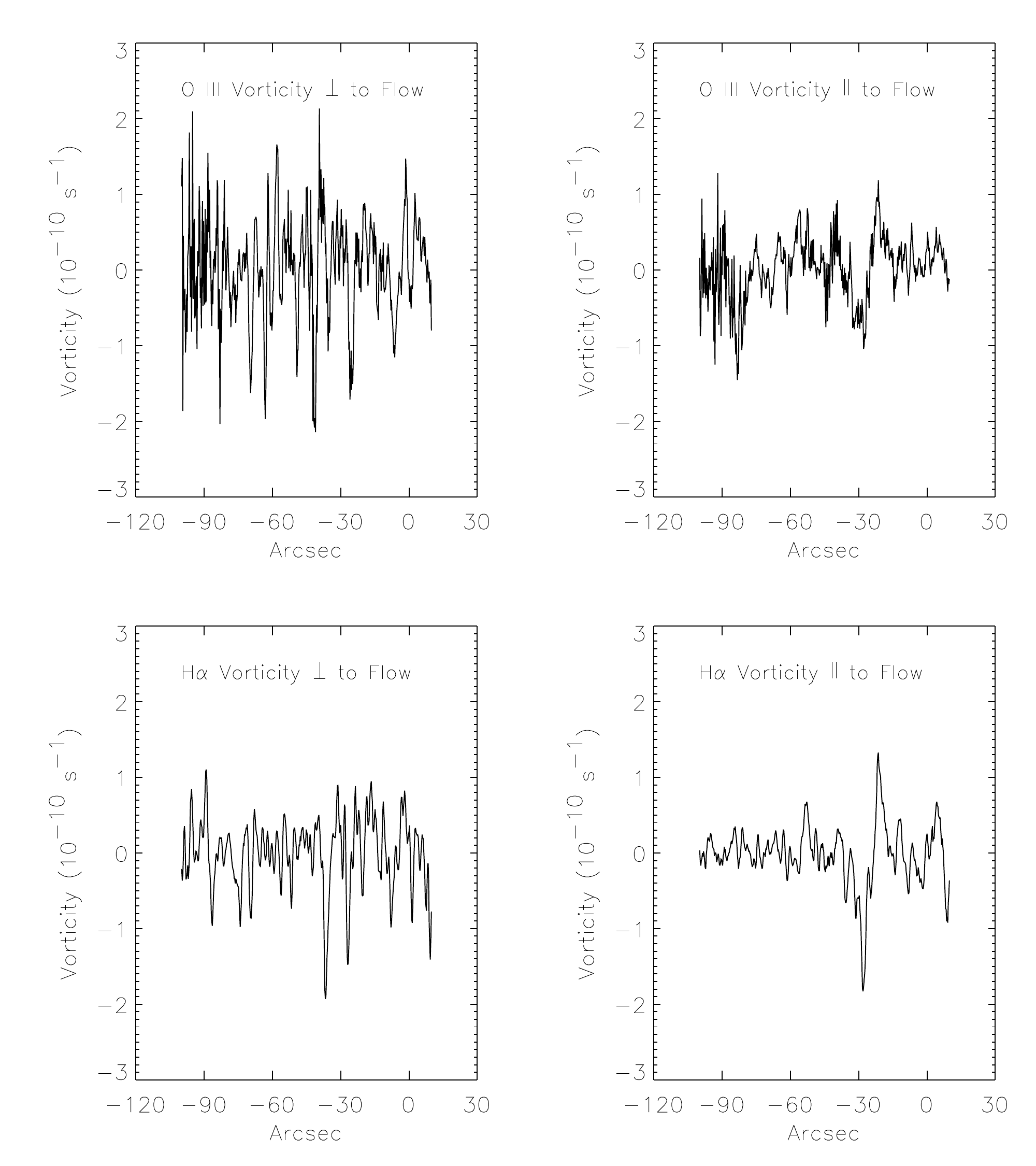}
\caption{Estimated vorticity as a function of position along the slit.  The component perpendicular to the flow corresponds to rotation around the line of sight.  It is therefore the component expected to wind up and amplify the magnetic field. Uncertainties in this estimate are discussed in the text.
\label{vorticity}
}
\end{figure*}

The resulting vorticities are shown in Figure~\ref{vorticity}.  As for the ram pressure estimate above, when the LOS intersects the shock at several places, some approaching and some receding from Earth, the velocities partially cancel and the derived vorticity is an underestimate.  We again argue that the highest vorticities found are in places where the LOS crosses one dominant position in the shock, so that the peak values are good estimates of the actual vorticity.  We also point out that comparison of the H$\alpha$ line width with observed velocity range indicates that there is little material with LOS velocities substantially larger than we measure.

The vorticity is the inverse of the turnover time.  That is roughly 170-300 yr, which is around one fourth the age of the radiative shock estimated from the spectrum.  That suggests that turbulence is partially developed and it could have begun to amplify the magnetic field.  It is worth noting that Figure~\ref{vorticity} shows that the vorticity perpendicular to the flow direction is larger than the component along the flow direction, indicating that turbulence has not developed far enough to become isotropic.  

It is somewhat surprising that the vorticity measured in H$\alpha$ is smaller than that measured in [O~III].  One possible explanation would be that the H$\alpha$ is more diffuse than [O~III] according to the 1D models, and that would tend to smear out gradients.  However, the images (Fig.~\ref{pm_ha}) show that the H$\alpha$ features are in fact sharper than those of [O~III].  It is possible that the turbulence damps out as the gas cools from [O~III] temperatures to H$\alpha$ temperatures.  That is especially plausible because of the dominance of magnetic pressure at the lower temperature.  It is also possible that the turbulence cascades to scales smaller than our resolution between the region where [O~III] forms and the H$\alpha$ emitting region.  Further discussion of the vorticity is deferred to a subsequent paper, which will use HST images to explore properties of the turbulence.

\section{DISCUSSION}

\begin{table}
\begin{center}
\caption {Summary of Shock Properties}
\begin{scriptsize}
\begin{tabular}{l c l l}
\hline
\hline
Property                 &  Value         & Units &  Basis         \\
\hline
$V_S$                    &  130           & km $s^{-1}$  & Proper motion, Dist.  \\
$P_{\mathrm{ram}} ~$&  $4 \times 10^{-9}$  & dyne $\rm cm^{-2}$  & [O III] Intens., V\\
$n_{\mathrm{pre}}$                &  6             & $\rm cm^{-3}$ & $P_{\mathrm{ram}}$ and V \\
$n_{\mathrm{post}}$               &  300           & $\rm cm^{-3}$ &  [S II] ratio,$n_e/n$ \\
X (Compression)          &  50            &   & $n_{\mathrm{post}} / n_{\mathrm{pre}}$  \\
$P_{\mathrm{th}}$    & $4.3\times 10^{-10}$   & dyne $\rm cm^{-2}$  & [S II] and [N II] \\
$B_{\mathrm{post}}$               & 290            & $\mu$G     & $P_{\mathrm{ram}} - P_{\mathrm{th}}$  \\
$B_{\mathrm{pre}}$                & 6              & $\mu$G     &  $B_{\mathrm{post}}$/X \\
$D_{\mathrm{LOS}}$                & 1.5$\times 10^{18}$ &  cm   & [S II] Intens., $n_e$ \\
$V_{\mathrm{turb}}$               &  20            & km $\rm s^{-1}$  & V variations \\
Vorticity                & 2$\times 10^{-10}$ & s$^{-1}$  & V gradients \\
Age                      & $\sim$ 1200    & yr & [O~III]/H$\alpha$ \\
$V_A$ before shock       & 5            & \kms  & $n_{\mathrm{pre}}$, $B_{\mathrm{pre}}$\\
$V_A$ at shock           & 10            & \kms  & $n_{\mathrm{pre}}$, $B_{\mathrm{pre}}$, X\\
$V_A$ [S II] zone        & 35             & \kms  & n$_{\mathrm{post}}$, $B_{\mathrm{post}}$ \\
$\beta$ at shock         & 68            &       & $n_{\mathrm{pre}}$, $B_{\mathrm{pre}}$, V \\
$\beta$ [S II] zone      & 0.13            &       & $n_{\mathrm{post}}$, $B_{\mathrm{post}}$, $T_{\mathrm{post}}$\\

\hline
\end{tabular}

\end{scriptsize}
\end{center}
\label{tab_shock}
\end{table}

The measured and derived properties of the shock are summarized in Table~\ref{tab_shock}.  Thanks to the combination of proper motions with an accurate distance to the Cygnus Loop, to the accuracy of emission line centroids from Binospec, and to the set of adjacent long slit spectra, it is possible to determine parameters such as ram pressure, magnetic field strength, compression ratio, and vorticity that are seldom available for interstellar shock waves.

Several important shock parameters are derived.  The combination of the ram pressure {\bf $\rho_0 V_S^2 = 4 \times 10^{-9}~\rm dyne~cm^{-2}$} with the shock speed of 130 \kms gives a preshock density of 6 $\rm cm^{-3}$.  Comparison of that pre-shock density with the electron density of 150-200 $\rm cm^{-3}$ in the [S~II] emitting region gives a total compression factor of 25-35.  However, the gas has begun to recombine by time it emits strongly in [S~II].   Based on the 1D models discussed in section 3.3, the gas is close to 50\% recombined in the region of strong [S~II] emission.  We therefore take the compression ratio to be 50.  Given the magnetic field estimate of 290 $\mu$G in that region and the compression factor, the perpendicular component of the preshock magnetic field is 6 $\mu$G.  The Alfv\'{e}n speeds in the preshock and cool postshock gas are 5 and 35 \kms, respectively.  The relative importance of magnetic forces is given by the plasma $\beta = P_{GAS}/P_{MAG}$.  Immediately behind the shock, assuming a shock compression factor of 4, $\beta$=68, while in the cool region downstream $\beta$=0.13.  This transition from high- to low-$\beta$ plasma as the gas cools will govern the development of turbulence in the flow.

\subsection{Energetic Particles}

Supernova remnant shock waves are believed to be a major source of cosmic-rays below the "knee" at around $10^{15}$ eV \citep{helder12}.  They can meet the energy requirements, and diffusive shock acceleration (DSA) naturally predicts approximately the observed power-law spectrum for the fast shock compression ratio of 4 \citep{blandfordeichler}.  Moreover, young SNRs are bright at radio, X-ray, and gamma-ray wavelengths due to the synchrotron, inverse Compton, and pion decay emission from the energetic particles they produce.  However, the efficiency of particle acceleration depends on the speed of the shock, and DSA is less effective in the slower shocks of older SNRs \citep{capriolispitkovsky14}. 

\citet{vanderlaan} proposed that the radio emission from middle-aged SNRs could be explained by the adiabatic compression of ambient cosmic-ray electrons and simple compressive strengthening of the magnetic field behind the shock.  The idea was applied to the SNRs IC443 and W44 by \citet{blandfordcowie} and \citet{cox99}, respectively.  More recently, acceleration by postshock turbulence and shock reacceleration \citep{zank15, caprioli18} have been proposed to enhance particle acceleration and modify the particle spectrum.  The densities and magnetic fields derived above provide a relatively clean test of the van der Laan mechanism.

To compute the radio emission we start with the electron spectrum given by \citet{potgieter15}, which matches the values measured outside the heliosphere by \citet{cummings16} with Voyager 1. We assume that the electrons are adiabatically compressed by a factor of 50 along with the gas, assuming $\gamma$=4/3, and we assume a compressed field strength of 290 $\mu$G (Table 2).  That gives an emissivity of {\bf $10^{-33.4}~\rm erg~cm^{-3}~s^{-1}~Hz^{-1}$} at 1 GHz.  \citet{green90} gives radio fluxes from this region, but the beam size of 6\farcm7 \/ by 4\farcm8\/  makes comparison difficult.  Instead we compare with the observations of bright radio filaments in the northeast Cygnus Loop by \citet{straka86} at 1.7 GHz with a 4\farcs7\/ beam.  To match their peak of 0.3 mJy per beam, we require an LOS depth of $2 \times 10^{17}~\rm cm$.  That is easily met by the highest LOS depths shown in Figure~\ref{dene_los} of order $10^{18}~\rm cm$.  We therefore conclude that the van der Laan mechanism can account for the observed radio emission from radiative shocks in the Cygnus Loop. 

The same procedure predicts synchrotron X-ray emission.  At 4 keV ($10^{18}$ Hz) the emissivity is $10^{-37.7}~\rm erg~cm^{-3}~s^{-1}~Hz^{-1}$, which implies a few times $10^{-5}$ photons $\rm cm^{-2} s^{-1}$ per 4\farcs7 spatial element with the bright filament LOS depth above.  Thus synchrotron X-ray filaments would be easily detectable with Chandra (effective area around 200 cm$^2$ at 5 keV) unless there is a break in the spectrum.  With a synchrotron cooling time of order 130 yr at 1 TeV, such a break is expected, and a high spatial resolution X-ray observation of the region might provide an interesting constraint.  \citet{katsuda08} included a power-law component in fits to the X-ray spectrum of part of the northeast Cygnus Loop and found a turnover frequency of $3 \times 10^{14}$ Hz.

The Cygnus Loop has also been detected in gamma-rays by Fermi \citep{katagiri11}.  They give a luminosity in GeV gamma rays of about $10^{33}~\rm erg~s^{-1}$, but at the 735 pc distance of \citet{fesen18} it would be $2 \times 10^{33}~\rm erg~s^{-1}$.  The gamma-rays are strongly correlated with the H$\alpha$ emission, which comes mainly from radiative shocks (their Figure 2), and a spectral break at 2--3 GeV suggests that the gamma-rays are produced by pion decay after energetic hadrons interact with dense gas.  Thus an origin in the strongly compressed region of radiative shocks is plausible.

The Voyager 1 measurements of energetic particles outside the heliosphere \citep{cummings16} show a flat spectrum in the 10-100 MeV range at about 30 protons $\rm m^{-2} ~s^{-1} ~sr^{-1} ~MeV^{-1}$.  If those protons are compressed by a factor of 50, adiabatically boosted in energy to around 1 GeV and placed in a thermal gas with a density of 300 $\rm cm^{-3}$ (twice the electron density shown in Figure~\ref{dene_los}), the pion decay emissivity is about 3$\times 10^{-24}~\rm ph~cm^{-3}~s^{-1}$.  The surface area of the Cygnus Loop is roughly 5500 square pc \citep{fesen18}, but the radiative shocks that provide strong compression only cover about six facets, making up perhaps 1/7 of the area \citep{hestercox}.  If we assume a shell thickness of $10^{17} \rm cm$ consistent with the Binospec observations and the thickness predicted by the shock models, the van der Laan mechanism implies a GeV gamma-ray luminosity of $10^{33}~\rm erg~s^{-1}$.  This is an order of magnitude prediction because of the large uncertainties in the shell thickness and covering factor of radiative shocks, and because the density and compression measured in one region may not be typical of radiative shocks in the rest of the remnant.  Moreover, the \citet{cummings16} particle flux implies a smaller cosmic-ray ionization rate than is generally inferred from observations of molecular clouds, so the ambient cosmic-ray flux near the Cygnus Loop could be significantly higher than was assumed.  Overall, the rough agreement of predicted and observed gamma-ray fluxes suggests that simple compression of pre-existing low-energy cosmic-ray protons in radiative shocks can account for the observed gamma rays.

The observations can also constrain models of magnetic field amplification and particle acceleration in collisionless shocks.  \citet{caprioli18} find that for moderate Mach numbers (M<30), the current due to streaming of ambient cosmic-rays reflected from the shock can drive the resonant Bell instability \citep{bell78}, which saturates at $\delta B/B ~ \sim$1. That implies regions where the shock is essentially parallel, permitting efficient injection of thermal particles into the acceleration process.  For our purposes, it also implies regions where the field lies along the shock normal and is therefore not compressed by the shock.  If those weak field regions survive, they will eventually be compressed into high-density knots when magnetic pressure begins to dominate \citep{raymondcuriel}.  However, \citet{caprioli18} found that they survive for a time corresponding to $~\sim~10^{10}$ cm, far below our resolution.  

\citet{caprioli18} also predict that faster shocks will amplify the magnetic field.  Our estimate of the preshock field based on the postshock field and the compression ratio is a typical value for parts of the ISM with densities of $\sim~5~\rm cm^{-3}$, indicating that there is little shock amplification of the field, in agreement with the predictions of \citet{caprioli18} for a Mach 13 shock.







\section{SUMMARY}

We have measured proper motions and spectra at many positions in a radiative shock in the Cygnus Loop.  The structure seen in low-temperature ions is far more complex than would be expected from 1D models of the cooling and photoionization regions behind radiative shocks. Thanks to the recently established distance to the Cygnus Loop \citep{fesen18}, the  proper motions provide a reliable shock speed.  From line intensities and Doppler shifts we determine the shock ram pressure, preshock and postshock densities and magnetic field strengths, and an estimate of vorticity (see Table 2).  

Comparison with radio and gamma-ray observations indicates that simple adiabatic compression of the ambient cosmic-rays and magnetic fields (van der Laan mechanism) can account for the nonthermal emission.  Reacceleration and magnetic field amplification by the shock \citep{caprioli18} do not seem to be required, as may be expected for the modest shock Mach number.

The average spectrum is matched surprisingly well by a 1D shock model, the major discrepancies being in the lowest ionization lines of [O I].  That suggests that unresolved shocks can be interpreted on the basis of grids of 1D models such as those of \citet{raymond79} or \citet{allen08} to estimate elemental abundances and shock parameters.  The presence of [Fe VI] and [Fe VII] lines at about one third of their predicted strengths suggests that dust grains are sputtered fairly efficiently in the hot region close to the shock.

A future paper will use the shock parameters established here to investigate the development of turbulence in the postshock cooling zone based on HST narrow-band images and spectra as the flow transitions from gas-dominated to magnetic-field-dominated regimes.  Higher spectral resolution observations and 3D magnetohydrodynamic simulations of the postshock cooling flow could greatly advance this investigation.

\acknowledgements
This work was supported by HST Guest Observer grant GO-15285.0001\_A to the Smithsonian Astrophysical Observatory. We thank the MMT staff for executing the observations in the queue mode. We thank Kirill Grishin at Moscow State University for providing his astrometry correction software package for testing prior to publication. I.C. is supported by Telescope Data Center at SAO and also acknowledges the Russian Science Foundation grant 19-12-00281 and the Program of development of M.V. Lomonosov Moscow State University for the Leading Scientific School ``Physics of stars, relativistic objects and galaxies.''. We obtained HST/WFPC2 images from the MAST archive at STScI. CHIANTI is a collaborative project involving George Mason University, the University of Michigan (USA), University of Cambridge (UK) and NASA Goddard Space Flight Center (USA). The analysis benefited from the Russian Academy of Sciences workshop on gamma-ray astronomy in 2019 September.

\facilities{MMT (Binospec)} {HST (WFPC2)} {HST (WFC3)}

\software{SWarp (Bertin et al. 2002)}

\bibliographystyle{aasjournal}
\bibliography{main}

\end{document}